\documentclass[useAMS,usenatbib,usegraphicx]{mn2e}
 
\title[Faint and Extremely Red K-band Selected Galaxies]{The Faint and Extremely Red K-band Selected Galaxy Population in the DEEP2/Palomar Fields}
\author[Conselice et al.]{C. J. Conselice$^{1,2}$\thanks{E-mail:
conselice@nottingham.ac.uk}, K. Bundy$^{2,3}$, Vivian U$^{2,4}$,  P. Eisenhardt$^{5}$, J. Lotz$^{6}$, \newauthor J. Newman$^{7}$ \\
 $^1$University of Nottingham, School of Physics \& Astronomy, Nottingham, NG7 2RD UK \\
 $^2$Previously: Palomar Observatory, Caltech, MC 105-24, Pasadena, CA \\
 $^3$Department of Astronomy, University of Toronto, Canada \\
 $^4$Institute for Astronomy, University of Hawaii \\
 $^5$Jet Proportion Laboratory, Caltech, Pasadena, CA \\
 $^6$NOAO, Tuscon, AZ \\
 $^7$Lawrence Berkeley National Laboratory, Berkeley CA}
\def\deg{$^{\circ}\,$}
\def\solm{M$_{\odot}\,$}

\def\deg{$^{\circ}\,$}
\def\solm{M$_{\odot}\,$}

\def\mass{$10^{11}$ M$_{\odot}\,$}
\def\hmass{$10^{11.5}$ M$_{\odot}\,$}

\begin{document}

\date{Accepted ; Received ; in original form}

\pagerange{\pageref{firstpage}--\pageref{lastpage}} \pubyear{2002}

\maketitle

\label{firstpage}

\begin{abstract}

We present in this paper an analysis of the faint and red near-infrared 
selected 
galaxy population found in near-infrared imaging from the
Palomar Observatory
Wide-Field Infrared Survey. This survey covers 1.53 deg$^{2}$ to 5 $\sigma$
detection limits of  K$_{\rm vega}$ = 20.5-21
and J$_{\rm vega}$ = 22.5, and overlaps with the DEEP2 spectroscopic  
redshift survey.   
  We discuss the details of this 
NIR survey, including our J and
K band counts. We show that the K-band galaxy population has a redshift
distribution that varies with K-magnitude, with most K $< 17$ galaxies
at $z < 1.5$ and a significant fraction (38.3$\pm0.3$\%) of $K > 19$ systems
at $z > 1.5$.  We further investigate the stellar masses and morphological 
properties of K-selected galaxies, particularly extremely red objects,
as defined by $(R-K) > 5.3$ and $(I-K) > 4$.  One of our conclusions is that
the ERO selection is a good method for picking out galaxies at $z > 1.2$,
and within our magnitude limits, the most massive galaxies at these 
redshifts.  The ERO limit finds 75\% of all M$_{*} > 10^{11}$ \solm
galaxies at $z \sim 1.5$ down to K$_{\rm vega} = 19.7$. We further find that 
the morphological break-down 
of $K < 19.7$ EROs is dominated by early-types (57$\pm3$\%) and 
peculiars (34$\pm3$\%).  However, about a fourth of the early-types
are distorted ellipticals, and within CAS parameter space these bridge the
early-type and peculiar population, suggesting a morphological evolutionary
sequence.  We also investigate the use of a $(I-K) > 4$ selection to
locate EROs, finding that it selects galaxies at slightly higher
average redshifts ($<z> = 1.43\pm0.32$) than the $(R-K) > 5.3$ limit
with $<z> = 1.28\pm0.23$.  Finally, by using the redshift distribution of 
$K < 20$ selected galaxies, and the properties of our EROs, we are able to 
rule out all monolithic collapse models for the formation of massive galaxies.

\end{abstract}


\section{Introduction}

Deep imaging and spectroscopic surveys in the optical have become the standard
method for determining the evolution of the galaxy population (e.g., Kron 1980;
Steidel \& Hamilton 1993; Williams et al. 1996; Giavalisco et al. 2004).
These surveys have revolutionised galaxy formation studies, and have allowed
us to characterise basic properties of galaxies, such as their luminosity 
functions, stellar masses, and morphologies, and how these properties have 
evolved (e.g., Lilly et al. 1995; Ellis 1997; Wolf et al. 2003; 
Conselice et al. 2005a).    However,
due to technological limitations with near-infrared arrays, most deep
surveys have been conducted in optical light, typically
between $\lambda =$ 4000-8000 \AA.  This puts limits on the usefulness of 
optical surveys, as they select galaxies in the rest-frame ultraviolet
at higher redshifts where many of the galaxies contributing to the
faint end of optical counts are located (e.g., Ellis et al. 1996).  Galaxies 
selected in the rest-frame ultraviolet limit our ability
to trace the evolution of the galaxy population in terms of stellar masses.
As the properties of galaxies can be quite different between the
rest-frame optical and UV (e.g., Windhorst et al. 2002; Papovich et al.
2003, 2005; Taylor-Manger et al.
2006; Ravindranath et al. 2006), it is desirable to trace galaxy evolution
at wavelengths where most of the stellar mass in galaxies is visible.  
To make advances in our understanding of galaxy evolution and formation at
high redshifts, ($z > 1$), therefore requires us to search for, and 
investigate, galaxy properties in the near-infrared (NIR).

Studying galaxies in the NIR has many advantages, including
minimised  K-corrections which are often substantial in the optical, 
as well as giving us a more direct probe of galaxy stellar mass up to
$z \sim 3$ (e.g.,
Cowie et al. 1994).  This has been recognised for many years, but most 
NIR surveys have been either deep 
pencil beam surveys (e.g., Gardner 1995; Djorgovski et al. 1995; Moustakas et al. 1997;
Bershady, Lowenthal \& Koo 1998;  Dickinson et al. 2000; Saracco et al. 2001;
Franx et al. 2003), or large-area, but shallow, surveys 
(e.g., Mobasher et al. 1986; Saracco et al. 1999;
McCracken et al. 2000; Huang et al. 2001; Martini 2001; Drory et al. 2001;
Elston et al. 2006).  This is potentially a problem for understanding
massive and evolved galaxies at high redshifts, as  red
objects are highly clustered (e.g., Daddi et al. 2000; Foucaud et al. 2007), 
as are the most massive galaxies (e.g., Coil et al. 2004a).  

Previous deep NIR surveys can detect these unique galaxy populations, 
but they typically do not have a large enough area to probe the range of
galaxies selected in the near-infrared. Likewise, large
area, but shallow surveys may not be deep enough to detect with a high
enough signal to noise these unique populations. 
In this paper we overcome this problem by presenting a  1.5 deg$^{2}$ 
survey down to 5 $\sigma$ depths of K$_{\rm vega}$ = 20.2-21.5 
and J$_{\rm vega} = 22.5$.
This  brings together the properties of both deep and wide surveys. 
In this paper we explore NIR galaxy counts, and study the 
properties of faint NIR galaxies, which are often red in 
near-infrared/optical colours.

Unique galaxy populations have long been known to exist in near-infrared 
selected surveys. 
These include the extremely red objects (Elston, Reike, Reike 1988) and the
distant red galaxies (Saracco et al. 2001; Franx et al. 2003; 
Conselice et al. 2007a; Foucaud et al. 2007), both of which are 
difficult to study at optical wavelengths.  The existence of these galaxies
reveals a large possible differential
in the galaxy population between optical and near-infrared surveys.  
While these objects can be located in deep optical surveys, they
are often very faint with $R > 26$ (e.g., van Dokkum et al.
2006), making it difficult 
to understand these objects in any detail without NIR imaging or
spectroscopy. In this paper we analyse the properties of galaxies selected in 
moderately deep K-band imaging.  We also investigate the redshift 
distributions, structures and properties of the near-infrared selected 
galaxy population down to $K_{\rm vega} \sim 20$.

One of our main conclusions are that the faint K-band population spans a range
of redshift and properties.  We find that galaxies with magnitudes
$K_{\rm vega} < 17$ are nearly all at $z < 1.4$.  Galaxies with 
magnitudes $K_{\rm vega} = 17-21$
are found at high redshift, up to at least $z \sim 4$.  The colours
of these galaxies span a wide range, with in particular redder galaxies seen
at higher redshifts.  Finally, we investigate the properties of the extremely
red objects in our sample, finding that they include most, but not all of,
the highest mass galaxies at $z \sim 1.5$.  The morphologies of these
EROs, and the redshift distribution and dust properties of $K_{\rm vega} 
< 20$ sources, show that hierarchical galaxy formation is the dominate 
method by which massive galaxies form.

This paper is organised as follows: in \S 2 we discuss the data
sources used in this paper, including our Palomar imaging, DEEP2 
redshifts, and HST ACS imaging. This section also gives basic details
of the Palomar survey.  \S 3 includes our analysis, which
contains information on the K and J-band counts, the redshift
and colour distributions of K-selected galaxies. \S 4 includes
an analysis of the extremely red galaxy population and its properties,
including stellar mass, dust content, and redshift distributions.  
\S 5 includes a
detailed discussion of our results in terms of galaxy models, while
\S 6 is a summary of our findings.  This paper uses
Vega magnitudes unless otherwise specified, and assumes a cosmology 
with H$_{0} = 70$~km s$^{-1}$ Mpc$^{-1}$, $\Omega_{\rm m} = 0.3$ and 
$\Omega_{\lambda} = 0.7$.

\section{Data, Reduction and Data Products}

\subsection{Data Sources}
 
The objects we study in this paper consist of those found in the 
fields covered by the Palomar Observatory Wide-Field Infrared Survey 
(POWIR, Table~1).  The  POWIR survey was designed to obtain deep K-band and 
J-band data over a significant ($\sim$1.5 deg$^2$) area.   Observations were 
carried out between September 2002 and October 2005 over a total of $\sim
70$ nights. This survey covers the GOODS field North (Giavalisco et al. 
2004; Bundy et al. 2005), the Extended Groth Strip (Davis et al. 2007), 
and three other fields the DEEP2 team
has observed with the DEIMOS spectrograph (Davis et al. 2003).  We however
do not analyse the GOODS-North data in this paper given its much smaller
area and deeper depth than the K-band imaging covering the DEEP2 fields.   
The total
area we cover in the K-band in the DEEP2 fields is 5524 arcmin$^{2}$ = 1.53 
deg$^{2}$, with half of this area imaged in the J-band. Our goal depth was
K$_{\rm vega} = 21$, although not all fields are covered this deep, but all 
have 5 $\sigma$ depths between K = 20.2-21.5.  Table~1 lists the DEEP2 
fields, and the area we have covered in each.  For our purposes we
abbreviate the fields covered as: EGS (Extended Groth Strip),
Field 2, Field 3, and Field 4.
 
The K-band data were acquired utilising the WIRC camera on the Palomar
5 meter telescope.  WIRC has an effective field of view of 
$8.1\arcmin \times 8.1\arcmin$, with a pixel scale
of 0.25\arcsec pixel$^{-1}$.  In total, our K-band survey consists of 75 WIRC 
pointings. During observations of the K data we used 30 second 
integrations with four exposures per pointing.  Longer exposure were 
utilised for the J-band
data, with an exposure time of 120 seconds per pointing.  
Total exposure times in both K and J were between one 
to eight hours. The seeing FWHM in the K-band data ranges from 0.8'' to 1.2'', 
and is on average 1.0''.  Photometric calibration was carried out by
referencing Persson standard stars during photometric 
conditions.   The final K-band and J-band images were made by combining 
individual mosaics  obtained over several nights. The K-band mosaics are 
comprised of co-additions of  $4 \times 30$ second exposures dithered over a 
non-repeating 7.0'' pattern. The J-band mosaics were analysed in a similar
way using single 120 second exposures per pointing. The images
were processed using a double-pass reduction pipeline we developed
specifically for WIRC.  For galaxy detection and photometry 
we utilised the SExtractor package (Bertin \& Arnouts 1996).

Photometric errors, and the K-band detection limit for each image were
estimated by randomly inserting fake objects of known magnitude into
each image, and then measuring photometry with the same detection parameters
used for real objects.  The inserted objects were given Gaussian
profiles with a FWHM of 1\farcs3 to approximate the shape of slightly
extended, distant galaxies. We also investigated the completeness and
retrievability of magnitudes for exponential and de Vaucouleurs profiles
of various sizes and magnitudes. A more detailed discussion of this is 
included in \S 2.4 and Conselice et al. (2007b).

Other data used in this paper consists of: optical
imaging from the CFHT over all of the fields, imaging from the Advanced Camera for 
Surveys (ACS) on the Hubble Space Telescope, and spectroscopy from the DEIMOS
spectrograph on the Keck II telescope (Davis et al. 2003).  A summary
of these ancillary data sets, which are mostly within the Extended Groth Strip,
are presented in Davis et al. (2007).

The optical data from the CFHT 3.6-m includes imaging in the B, R and I 
bands taken with the CFH12K camera, which is a 12,288 $\times$ 8,192
pixel CCD mosaic with a pixel scale of 0.21\arcsec.
The integration time for these observations are 1 hour in $B$ and $R$ and
2 hours in $I$, per pointing with 5 $\sigma$ depths of $\sim$ 25 in each 
band.   For details of the data reduction see
Coil et al. (2004b).  From this imaging
data a R$_{\rm AB}$ = 24.1 magnitude limit was used for determining targets 
for 
the DEEP2 spectroscopy.   The details for how these imaging data were acquired 
and reduced, see Coil et al. (2004b).
  
The Keck spectra were obtained with the DEIMOS spectrograph 
(Faber et al. 2003) as part of the DEEP2 redshift survey. 
The EGS spectroscopic sample was selected based on a R-band
magnitude limit only (with $R_{\rm AB} < 24.1$), with no strong colour cuts 
applied to the 
selection.  Objects in Fields 2-4 were selected for spectroscopy based on 
their position in $(B-K)$ vs. $(R-I)$ colour space, to locate galaxies 
at redshifts $z > 0.7$.   The total DEEP2 survey includes over 30,000 galaxies 
with a secure redshift, with about a third of these in the EGS field, and in
total $\sim 11,000$ with a K-band detection (\S 3.1.1).  In 
all fields the sampling rate for galaxies that meet the selection criteria is
60\%.

The DEIMOS spectroscopy was obtained using the
1200 line/mm grating, with a resolution R $\sim 5000$ covering
the wavelength range 6500 - 9100 \AA.  Redshifts were measured through
an automatic method comparing templates to data, and we only utilise
those redshifts measured when two or more lines were
identified, providing very secure redshift measurements.  Roughly 70\% of all
targeted objects resulted in reliably measured redshifts.  Many of the redshift
failures are galaxies at higher redshift, $z > 1.5$ (Steidel et al.
2004), where the [OII] $\lambda$3727 lines leaves the optical window. 

The ACS imaging over the EGS field covers a 10.1\arcmin $\times$ 70.5\arcmin\,
strip, for a coverage area of 0.2 deg$^{2}$.  The ACS imaging
is discussed in Lotz et al. (2006), and is briefly described
here, and in Conselice et al. (2007a,b). The imaging consists of 63 titles 
imaged in both the F606W (V) and F814W (I) bands.  The 5-$\sigma$ depths 
reached in these images are V = 26.23 (AB) and I = 27.52 (AB) for a point
source, and about two magnitudes brighter for extended objects.

Our matching procedures for these catalogs progressed in the manner 
described in Bundy et al. (2006). The K-band 
catalog served as our
reference catalog. We then matched the optical catalogs and spectroscopic
catalogs to this, after correcting for any astrometry issues by
referencing all systems to 2MASS stars.   All magnitudes quoted
in this paper are total magnitudes, while colours are measured
through aperture magnitudes.

\setcounter{table}{0}
\begin{table}
 \caption{The Palomar Fields, Number of WIRC pointings, and Area Covered}
 \label{tab1}
 \begin{tabular}{@{}lccccc}
  \hline
Field & RA & Dec. & \# K & \# J & K-area (arcmin$^{2}$) \\
\hline
EGS & 14 17 00 & +52 30 00 & 33 & 10 & 2165 \\
Field 2 & 16 52 00 & +34 55 00 & 12 & 0 & 787 \\
Field 3 & 23 30 00 & +00 00 00 & 15 & 15 & 984 \\
Field 4 & 02 30 00 & +00 00 00 & 15 & 12 & 984 \\
\\
Total   &          &           & 75 & 37 & 4920 \\
\hline
 \end{tabular}
\end{table}

\subsection{Photometric Redshifts}

We calculate photometric redshifts for our K-selected galaxies, which do
not have DEEP2 spectroscopy, in a number of ways. This sample is hereafter
referred to as the `phot-z' sample.  Table~2 lists the number of
spectroscopic and photometric redshifts within each of our K-band magnitude
limits.    These photometric redshifts are based on
the optical+near infrared imaging, BRIJK (or BRIK for half the data) 
bands, and are fit in two ways, depending on the brightness of a galaxy in 
the optical.   For galaxies that
meet the spectroscopic criteria, $R_{\rm AB} < 24.1$, we utilise a neural
network photometric redshift technique to take advantage of the
vast number of secure redshifts with similar photometric data.  Most
of the $R_{\rm AB} < 24.1$ sources not targeted for spectroscopy should be within our
redshift range of interest at $z < 1.4$.    The neural network fitting is 
done through 
the use of the ANNz (Collister \& Lahav 2004) method and code.
To train the code, we use the $\sim 5000$ redshifts in the EGS, which
span our entire redshift range.  The agreement between our 
photometric redshifts and our ANNz spectroscopic redshifts is very good 
using this technique, with $\delta z/z = 0.07$ out
to $z \sim 1.4$. The photometry we use for our photometric
redshift measurements are done with a 2\arcsec\, diameter aperture.

For galaxies which are fainter than $R_{\rm AB} = 24.1$ we utilise photometric
redshifts using two methods, depending on whether the galaxy is
detected in all optical bands or not.  For systems which are detected
at all wavelengths we use the Bayesian approach from
Benitez (2000).  For an object to have a photometric redshift using
this method requires it to be detected at the 3 $\sigma$ level in all
optical and near-infrared (BRIJK) bands, which in the R-band
reaches $\sim 25.1$. We refer to these objects as having `full' photometric
redshifts.  As described in Bundy et al. (2006) we 
optimised our results and corrected for systematics through 
the comparison with spectroscopic redshifts, resulting
in a redshift accuracy of $\delta z/z = 0.17$ for $R_{\rm AB} > 24.1$ systems. 
Further details about our photometric redshifts are presented in
Conselice et al. (2007b), including a lengthy discussion of
biases that are potentially present in the measured values.

Table~2 lists the number of galaxies with the various redshift types.
As can be seen, the vast majority of our galaxies have either spectroscopic
redshifts, or have measured photometric redshifts using the full optical
SED.  Only a small fraction ($< 1$\%) of our K-band sources are not
detected in one optical band down to $K = 21$.    

For completeness in
the analysis of the $N(z)$ distribution of K-magnitudes discussed in \S 5,
we calculate, using a $\chi^{2}$ minimisation through hyper-z (Bolzonella,
Miralles \& Pello 2000), the
best fitting photometric redshifts for these faint systems.  These
galaxies however make up only a small fraction of the total
K-band population, and their detailed redshift distribution, while
not likely as accurate as our other photometric redshifts, do not
influence the results at all significantly.

\subsection{Stellar Masses}

From our K-band/optical catalogs we compute stellar masses based on the methods
outlined in Bundy, Ellis, Conselice (2005) and Bundy et al. (2006). The
basic method consists of fitting a grid of model SEDs constructed
from Bruzual \& Charlot (2003) stellar population synthesis models, with
different star formation histories.  We use an exponentially declining model
to characterise the star formation history, with various ages, 
metallicities and dust contents included.  These models are parameterised
by an age, and an e-folding time for characterising the star formation 
history.   We also investigated how these stellar masses would change
using the newest stellar population synthesis models with the latest
prescriptions for AGB stars from Bruzual \& Charlot (2007, in prep). We
found stellar masses that were only slightly less, by 0.07 dex, compared
to the earlier models (see Conselice et al. 2007b for a detailed 
discussion of this and other stellar mass issues.)

Typical random errors for our stellar masses are 0.2 dex from the width
of the probability distributions.  There are also uncertainties from
the choice of the IMF.  Our stellar masses utilise the
Chabrier (2003) IMF, which can be converted to Salpeter IMF stellar masses 
by adding 0.25 dex.  There are additional
random uncertainties due to photometric errors.  The resulting
stellar masses thus have a total random error of 0.2-0.3 dex,
roughly a factor of two.  However, using our method we find that stellar
masses are roughly 10\% of galaxy total masses at 
$z \sim 1$, showing their reliability (Conselice et al. 2005b). 
Details on the stellar masses we utilise,
and how they are computed, are presented in Bundy et al. (2006) and
Conselice et al. (2007b).

\subsection{K-band Completeness Limit}

Before we determine the properties of our K-selected
galaxies,  it is first important to characterise how our detection
methods, and reduction procedures, influence the production of the 
final K-band catalog.
While the major question we address is the nature of the faint and red galaxy
population, it is important to understand what fraction of the faint
population we are missing due to incompleteness.  To understand this 
we investigate the K-band completeness of our sample in a number of ways.
The first is through simulated detections of objects in our
near-infrared imaging.  As described in Bundy et al.
(2006), Conselice et al. (2007b), and Trujillo et al. (2007) 
we placed artificial objects into our K-band images to
determine how well we can retrieve and measure photometry for galaxies
at a given magnitude.
Our first simulations were performed from K = 18 to K = 22 using
Gaussian profiles.  We
find that the completeness within our fields remains high at nearly all
magnitudes, with a completeness of nearly
100\% up to $K = 19.5$ for all 75 fields combined together.  The
average completeness of these fields at $K = 20$ is 94\%, which
drops to 70\% at $K = 20.5$ and 35\% at $K = 21.0$.

If we take the 23 deepest fields 
we find a  completeness at $K = 21.0$ of 70\%. However, galaxies
are unlikely to have Gaussian light profiles, and as such, we
investigate how the completeness would change in Conselice et al.
(2007b), if our
simulations were carried out with exponential and r$^{1/4}$ light
profiles. We find similar results as when using the Gaussian profiles up to
$K = 20$, but are less likely to detect faint galaxies with 
r$^{1/4}$ profiles, and
retrieve their total light output.  As discussed in \S 3.1.1, these
incompleteness corrections are critical for obtaining accurate galaxy
counts, but the intrinsic profiles of galaxies of interest must be
known to carry this out properly.  As such, we utilise the
Gaussian corrections as a fiducial estimate.   In Figure~1 we
plot our K-band counts with these corrections applied. We also
plot the J-band counts up to their completeness limit, and do not
apply any corrections for incompleteness.
The 100\% completeness for the optical data 
is $B = 25.25$, $R = 24.75$, and $I = 24.25$ (Coil et al. 2004b). 
These limits are discussed in \S 4.1 where we consider our
ability to retrieve a well defined population of extremely red objects (EROs).

\section{Analysis}

\subsection{Nature of the Faint K-band Population}

\subsubsection{K-band, J-band Counts and Incompleteness}

%numbers updated Jan. 27th, 2007

Within our total K-band survey area of 1.53 deg$^{2}$ 
we detect 61,489 sources at all magnitudes,
after removing false artifacts.  Most of these objects (92\%) are at K $< 21$,
while 68\% are at K $< 20$ and 37\% are at K $< 19$. In total there
are 38,613 objects fainter than K $= 19$ in our sample.  
Out of our total K-band population 10,693 objects (mostly galaxies) have 
secure spectroscopic DEIMOS redshifts from the DEEP2 redshift survey (Davis et 
al. 2003).  We supplement these by 37,644 photometric redshifts within
the range $0 < z < 2$ (Table~2).  We remove stars from our catalogs, 
detected through their structures and colours, as described in Coil 
et al. (2005) and Conselice et al. (2007b).

\setcounter{table}{1}
\begin{table}
 \caption{Number of K-band selected galaxies with various redshift measurements}
 \label{tab1}
 \begin{tabular}{@{}lcccc}
  \hline
K Range & Spec-z & Full Photo-z & Photo-z & Total \\
\hline
$15 < K < 17$ & 353 & 1541 & 1 & 1895 \\
$17 < K < 19$ & 4305 & 11379 & 30 & 15714 \\
$19 < K < 21$ & 5483 & 24405 & 215 & 30103 \\
\hline
 \end{tabular}
\end{table}

We plot the differential number counts (Table~3) 
for galaxies in both the K and J-band for our
K-selected sample in Figure~1 to test how our counts compare with those found
in
previous deep and wide near-infrared surveys.  We carry this out to determine
the reliability of our star and galaxy separation methods, as well
as for determining how our incompleteness
corrections in the K-band compare to others. As Figure~1 shows, we find little 
difference in our counts compared to previous surveys,
and we are $\sim$100\% complete up to magnitude K$\sim 20$ in all fields. 
As others have noted, we find a change in the slope of the galaxy
counts at K = 17.5.  We calculate that the slope at $K < 17.5$ is dN/dK = 
$0.54\pm0.07$, while at $K > 17.5$ it is dN/dK = $0.26\pm0.01$.   

Our counts, after correction,
are slightly lower than those in the UKIDSS UDS survey (Simpson et al.
2006), and from studies by Cristobal-Hornillos et al. (2003) and 
Saracco et al. (1999).  However, our counts are higher than those found in
Iovino et al. (2005) and Kong et al. (2006).   At brighter magnitudes these
surveys all agree, with the exception of Maihara et al. (2001) who
underpredict all surveys (not plotted). 
This difference at $K > 20$ is likely the result of
the different incompleteness correction methods used.  As detailed
in Bershady et al. (1998), using various intrinsic galaxy profiles when
computing completeness can lead to over
and underestimation of the correction factor.  The only
accurate way to determine the incompleteness is to know the
detailed distribution of galaxy surface brightness profiles at 
the magnitude limits probed (Bershady et al. 1998). As this is difficult,
and sometimes impossible to know, all corrected counts must
be seen as best estimates.

% & Error$_{\rm upper}$ & Error$_{\rm lower}$
\setcounter{table}{2}
\begin{table*}
 \caption{K-band and J-band Counts for all Fields}
 \label{tab1}
 \begin{tabular}{@{}cccc}
  \hline
K-Magnitude & log N (deg$^{-2}$ mag$^{-1}$) & J-Magnitude & log N (deg$^{-2}$ mag$^{-1}$)\\
\hline
          14.0   &    0.969$^{+0.139}_{-0.206}$ & 14.5 &  1.074$^{+0.176}_{-0.301}$ \\
        14.5   &    1.270$^{+0.102}_{-0.135}$ & 15.0 &  0.949$^{+0.198}_{-0.374}$ \\
          15.0   &    1.987$^{+0.048}_{-0.054}$ & 15.5 &  0.949$^{+0.198}_{-0.374}$ \\
        15.5   &    2.349$^{+0.032}_{-0.035}$ & 16.0 &  1.852$^{+0.081}_{-0.010}$ \\
          16.0   &    2.697$^{+0.022}_{-0.023}$ & 16.5 &  2.265$^{+0.052}_{-0.059}$ \\
        16.5   &    3.000$^{+0.016}_{-0.016}$ & 17.0 &  2.540$^{+0.038}_{-0.042}$ \\
          17.0   &    3.291$^{+0.011}_{-0.011}$ & 17.5 &  2.847$^{+0.027}_{-0.029}$ \\
        17.5   &    3.517$^{+0.009}_{-0.009}$ & 18.0 &  3.152$^{+0.019}_{-0.020}$ \\
          18.0   &    3.720$^{+0.007}_{-0.007}$ & 18.5 &  3.370$^{+0.015}_{-0.016}$ \\
        18.5   &    3.889$^{+0.006}_{-0.006}$ & 19.0 &  3.631$^{+0.011}_{-0.012}$ \\
          19.0   &    4.029$^{+0.005}_{-0.005}$ & 19.5 &  3.840$^{+0.008}_{-0.009}$ \\
        19.5   &    4.171$^{+0.004}_{-0.004}$ & 20.0 &  4.004$^{+0.007}_{-0.008}$ \\
          20.0   &    4.314$^{+0.003}_{-0.004}$ & 20.5 &  4.185$^{+0.006}_{-0.006}$ \\
        20.5   &    4.396$^{+0.003}_{-0.003}$ & 21.0 &  4.288$^{+0.005}_{-0.005}$ \\
          21.0   &    4.482$^{+0.003}_{-0.003}$ & 21.5 &  4.315$^{+0.005}_{-0.005}$ \\
\hline
 \end{tabular}
\end{table*}

The J-band counts (Figure~1b) show a similar pattern as the
K-band counts.  These counts are not corrected for incompleteness
and we are incomplete for very blue galaxies with $(J-K) < 0$
at the faintest J-band limit due to our using the K-band detections
as the basis for measuring J-band magnitudes. As can be seen, 
there is a larger variation in the J-band number counts when comparing
the different surveys at a given magnitude compared to the K-band
counts.  Furthermore, there is no obvious slope change in the J-band
counts, as seen in the K-band.   We are complete overall in 
the J-band to J$_{\rm vega} = 22.0$ over the entire survey. Our
J-band depth however varies between the three fields in which
we have J-band coverage, and in fact varies between individual WIRC
pointings. 

As we later discuss the properties of EROs in this paper, as defined
with a $(R-K)$ colour cut, it is
important to understand the corresponding depths of the
$R$-band imaging. The depth and number counts for
the $R$-band imaging is discussed in detail in Coil et al. (2004b).
Based on aperture magnitudes the 5 $\sigma$ depth of the CFHT R-band imaging
is roughly $R_{\rm AB} = 25$.  Our
$R$-band photometry uses the same imaging as Coil et al. (2004b),
however we retrieved our own magnitudes based on the $K$-band
selected objects in our survey. Our $R$-band depth however
is similar to Coil et al. (2004b), and we calculate a 50\%
incompleteness at $R = 25.1$.

%numdk in ero.sm /anal8/
\begin{figure*}
%\vspace{5.5cm}
 \vbox to 150mm{
\includegraphics[angle=0, width=154mm]{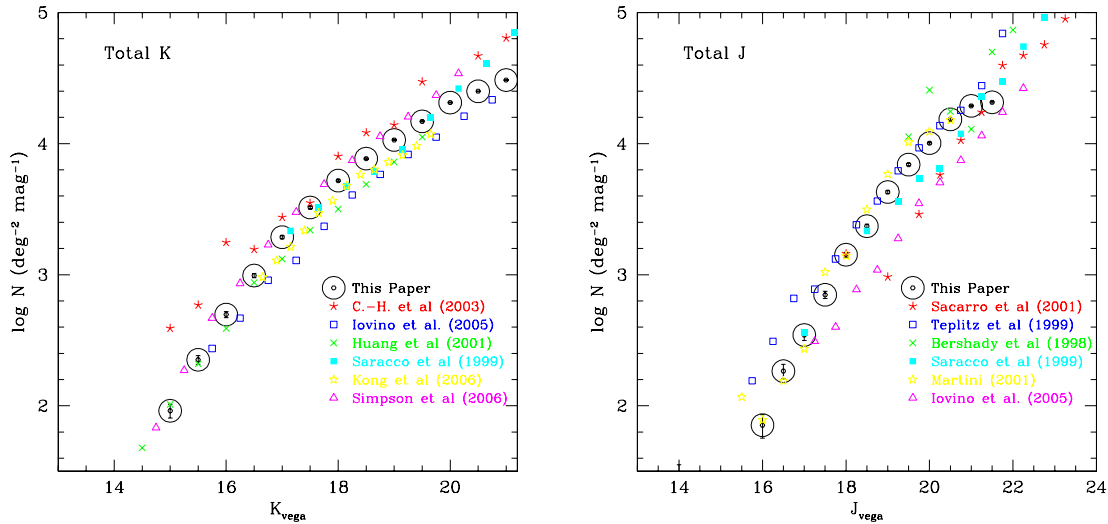}
%\vspace{-0cm}
 \caption{Differential K-band and J-band counts for our survey compared to 
previous published  counts, including: 
Cristobal-Hornillos et al. (2003) (C.-H.), Huang et al. (2001), 
Saracco et al. (1999), Saracco
et al. (2001), Teplitz et al. (1999), Bershady et al. (1998), Martini (2001), 
Kong et al. (2006), Iovino et al. (2005) and Simpson et al. (2006).   Only the error bars
are shown for the counts in our survey.  The K-band counts at $K > 20$
have been corrected for incompleteness, while the J-band counts are
only shown to their completeness limit.   At the faintest magnitudes 
the other surveys, with the exception of Simpson et al. (2006), have a 
larger error range due to the smaller areas used. }
} \label{sample-figure}
\vspace{-4cm}
\end{figure*} 

\subsubsection{Redshift Distributions of K-Selected Galaxies}

In this section we investigate the nature of galaxies selected
in the K-band. The basic question we
address is what are the properties of galaxies at various K-limits.
This issue has been discussed earlier by Cimatti et al. (2002a),
Somerville et al. (2004) and others. However, we are able
to utilise the DEEP2 spectroscopic survey of these fields to determine
the contribution of lower redshift galaxies to the K-band counts, and
thus put limits on the contribution of high redshift $(z > 1.5)$ 
galaxies to K-band counts at $K < 20$.

%ero.sm, kvz
\begin{figure}
%\vspace{5.5cm}
 \vbox to 80mm{
\includegraphics[angle=0, width=84mm]{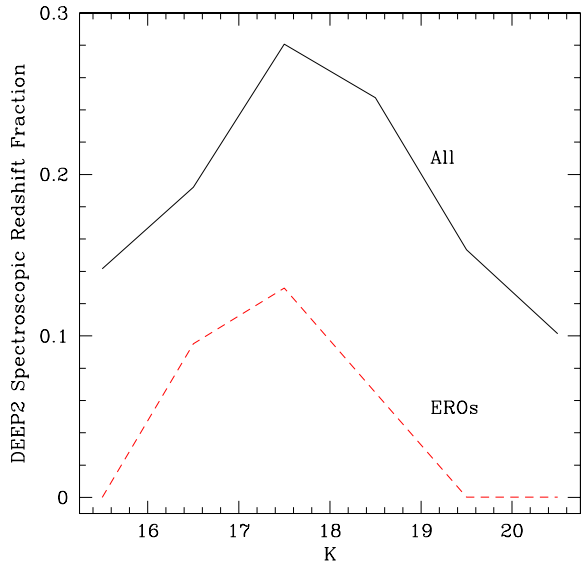}
%\vspace{-0cm}
 \caption{The spectroscopic redshift completeness for both the
entire K-band selected catalog, and for the
the $(R-K) > 5.3$ ERO selected sample. Note that none of our
EROs at $K > 19.5$ have a measured spectroscopic 
redshift, typical for spectroscopic surveys which
are optically selected.}
} \label{sample-figure}
\vspace{2cm}
\end{figure}  

%ero.sm, redd
\begin{figure}
\hspace{-0cm}
 \vbox to 80mm{
\includegraphics[angle=0, width=84mm]{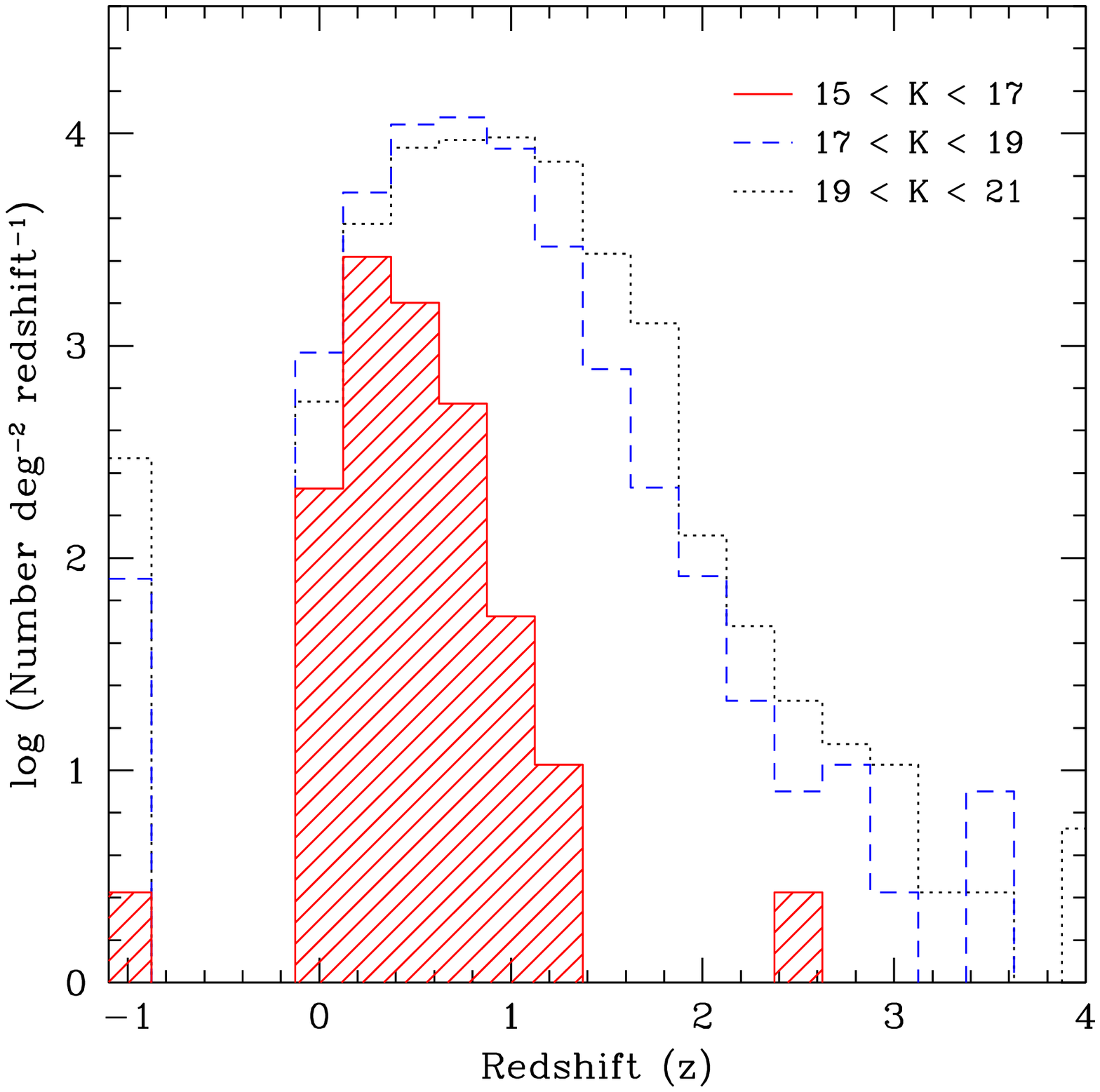}
%\vspace{-0cm}
 \caption{Redshift distributions for various K-band
magnitude cuts.  The solid black line shows the
redshift distribution for all galaxies with $19 < K < 21$,
the blued dashed line shows systems with $17 < K < 19$,
and the red hashed histogram is for galaxies
with $15 < K < 17$.  The levels at $z = -1$ show the
number of galaxies who had photo-zs  not fit due to
lacking significant optical detections.   }
} \label{sample-figure}
\vspace{3cm}
\end{figure}

The first, and most basic, method for understanding galaxies found in
a K-band
selection is to determine what fraction of the K-selected galaxies
have a successfully measured spectroscopic redshift. 
The  DEEP2 spectroscopic redshift success rate for our K-selected 
sample varies
with K-band magnitude, from 10\% to 30\%, up to $K = 21$. The highest
selection fraction is 30\% at K = 17.5. At the faintest limit, K = 21,
the redshift selection is 10\%, and the fraction is 15\%
at K = 15.5.  Figure~2 shows our spectroscopic redshift completeness 
as a function of
K-band magnitude for both the entire K-band selected sample,
and the EROs (\S 4).  The result of this plot is partially due to
the fact that the DEEP2 selection deweights galaxies at
$z < 0.7$.   The EGS and the other fields also have slightly
different methods for choosing redshift targets (Davis et al. 2003), 
creating an  inhomogeneous selection over the entire survey.

When we include photometric redshifts to supplement our spectroscopic 
redshifts, we
obtain total redshift distributions shown in Figure~3. Note that our
photometric redshifts are only included in Figure~3 at $z > 0$ if the object
was significantly detected in all bands in the BRIJK photometry.  In
each K-band limit shown in Figure~3, and discussed below, there are
a fraction of sources which do not meet this optical criteria, and these
objects are counted at the $z = -1$ position on the redshift histograms.  
In Table~2 we list the
number of K-band detections with and without spectroscopic redshifts, in
each of the redshift ranges.  

It appears that nearly all bright K-band sources, with $15 < K < 17$, 
are located at $z < 1.4$ (Figure~3). At fainter magnitudes, as shown
by the plotted $17 < K < 19$ and $19 < K < 21$ ranges (Figure~3), we
find a different distribution skewed toward high redshifts. While there are 
low redshift galaxies
at these fainter K-limits, we also find a significant contribution
of sources at higher redshifts, including those at $z > 2$.  The
K-bright sources at these redshifts are potentially the highest
mass galaxies in the early universe.  Galaxies at the faintest magnitudes, at
$19 < K < 21$, show a similar redshift distribution as the 
galaxies within the $17 < K < 19$ magnitude range, but there are a larger 
number of higher redshift galaxies. This demonstrates that 
faint K-band sources are just as likely to be at low redshift as at
high redshift. 

This is shown in another way using Figure~4 where we plot the distribution
of K-magnitude vs. redshift ($z$).  As can be seen, at $K > 19$ the
entire redshift range is sampled, while a $K < 17$ selection only finds
galaxies at $z < 1.4$. 

%ero.sm, kvz
\begin{figure}
%\vspace{5.5cm}
 \vbox to 80mm{
\includegraphics[angle=0, width=84mm]{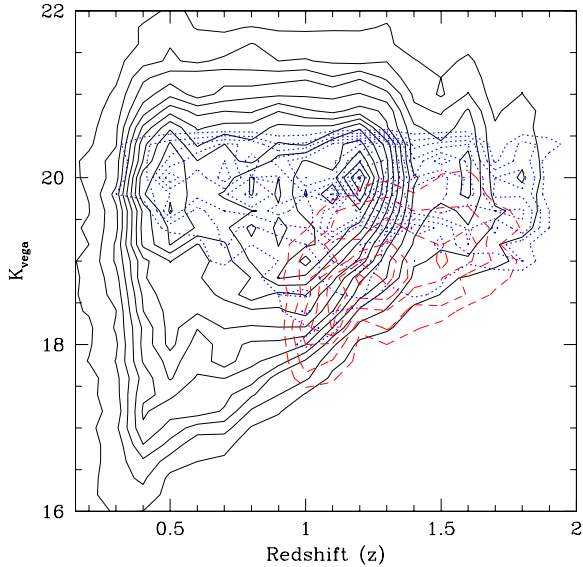}
%\vspace{-0cm}
 \caption{Contours of the redshift distribution for
galaxies at various K-magnitudes.  The red dashed
contours are for EROs defined as (R-K)$>5.3$ (mostly
at $z > 1$), and
the blue dotted contours are the distant red galaxies (DRGs) defined
by $(J-K) > 2.3$, which generally span $0.4 < z < 2$.}
} \label{sample-figure}
\vspace{4cm}
\end{figure}  

\subsubsection{Colours of K-Selected Galaxies}

After examining the redshift distribution of our sample,
the next step is determining the physical features of these galaxies.
The easiest, and most traditional, way to do this is through
the examination of colour-magnitude diagrams.  Generally,
galaxy colour is a mixture of at least three effects - redshift,
stellar populations and dust.  Galaxies generally become redder
with redshift due to band-shifting effects, and become redder
with age, and increased dust content.   

\begin{figure*}
%\vspace{5.5cm}
 \vbox to 150mm{
\includegraphics[angle=0, width=154mm]{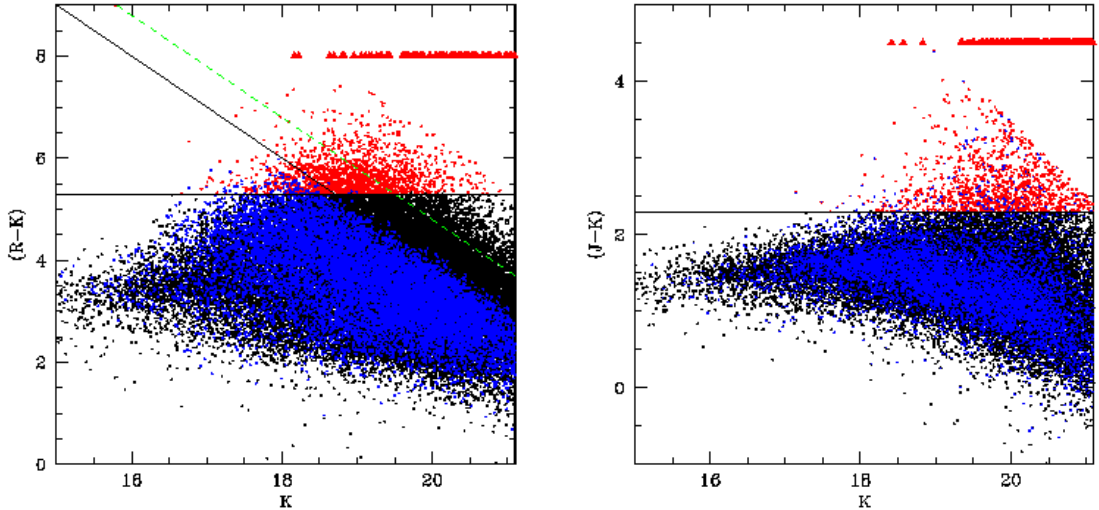}
%\vspace{-0cm}
 \caption{Colour-magnitude diagrams for our sample.  The left
panel shows (R-K) colour vs. K-band magnitude, while the right
panel shows the (J-K) vs. K diagram.  Objects with spectroscopic
redshifts are coloured blue in both diagrams, while objects
considered `red' either through the extremely red objects (EROs) or
Distant Red Galaxies (DRGs) selection are labelled as red in each diagram.
The solid line in the (R-K) vs. K panel shows the spectroscopic
limit of R = 24.1, while the dashed line shows the 5$\sigma$ 
limit for the R-band photometry of 25.1.  Furthermore, we only
plot points that are brighter than R = 26.5 and J = 23.5 in the
two panels, respectively.  The red triangles at the top of each
figure are galaxies which are undetected in R or J, but have
a measured K-band magnitude. }
} \label{sample-figure}
\vspace{-4.5cm}
\end{figure*} 

We can get an idea of the characteristics of  our K-selected sample
by examining the colour-magnitude diagram for the entire
$K < 21$ sample (Figure~5).  Figure~5 plots the colours
of our sample, as a function of $(B-R)$ and $(J-K)$ versus
K-band magnitude.  As can be seen, at fainter
limits there are more red galaxies in each band. Since
fainter/redder galaxies are more likely than brighter galaxies
to be at higher redshifts, it is likely that these
redder galaxies seen in Figure~5 are distant galaxies.  The relation
between $(R-K)$ and redshifts 
(Figure~6) shows this to be the case.  As can be seen, at
higher redshifts galaxies are redder in $(R-K)$, although
even at these redshifts there are K-selected galaxies
which are very blue.  At the highest redshifts, where optical
magnitudes are at $R > 24$, we find that
most galaxies hover around $(R-K) = 5.3$. However, as
can be seen, a significant fraction of the $K < 19.7$ galaxies,
which are massive systems at $z > 1$, would be identified as
EROs.

\begin{figure*}
%\vspace{5.5cm}
 \vbox to 150mm{
\includegraphics[angle=0, width=154mm]{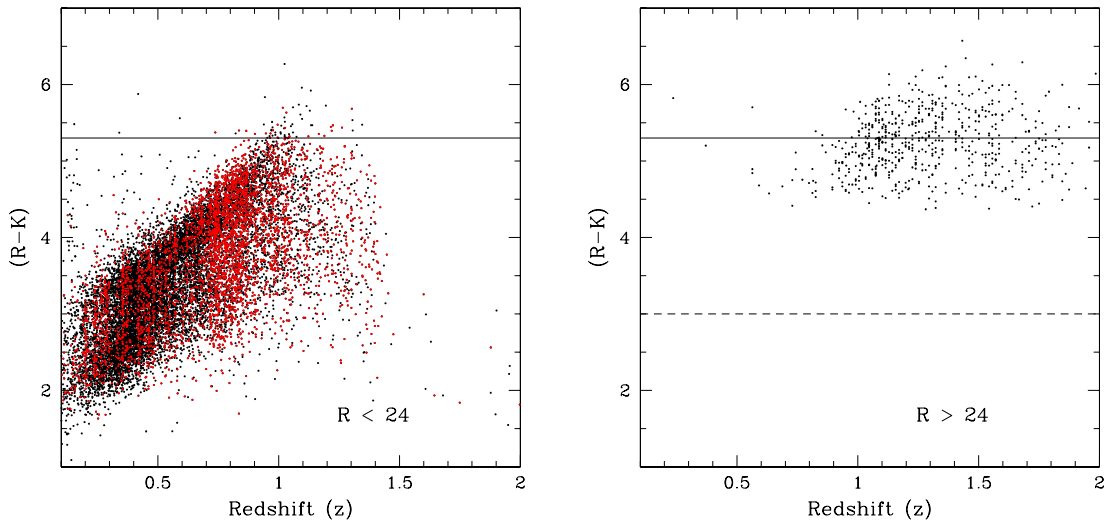}
 \caption{The redshift distribution for galaxies within
our sample at $K < 19.7$.  The left panel shows systems which are
at $R < 24$.  The right
panel shows the distribution of $(R-K)$ colours as a
function of redshift for galaxies which are fainter
than $R = 24$. As can be seen, galaxies generally get
redder at higher redshifts, but there still exists
a scatter in colour at any redshift.}
} \label{sample-figure}
\vspace{-6cm}
\end{figure*}

The detailed distribution of magnitudes, colours and stellar
masses is shown in Figure~7 and Figure~8, divided into different
redshift bins. Plotted
with different symbols are the photometric redshift and the
spectroscopic redshift samples.  As can been seen, there are
strong relations between stellar mass and K-band magnitude
over the entire redshift range (Figure~7), with fainter K-selected
galaxies having lower stellar masses.  Also note that the galaxies 
with spectroscopic redshifts are generally brighter and bluer than the 
photometric redshift sample at a given stellar mass.    
This is particularly true
at the highest redshift bins, and demonstrates that the
spectroscopy is successfully measuring the brighter galaxies
in the distant universe, while being less efficient in
measuring redshifts for galaxies of the same mass, but at
a fainter K-magnitude.  

Figure~8 furthermore shows how, as we go to higher redshifts,
we obtain an overall redder distribution of colours.  Within our
lowest redshift bin, $0.5 < z < 0.75$, there are few
galaxies which would be classified as EROs with $(R-K) > 5.3$.
It is also at this lowest redshift range where the overlap
between the spectroscopic and photometric samples is highest.
When we go to higher redshifts, such as at $0.75 < z < 1.0$,
we find that the slope of the locus in the relation between
stellar mass and $(R-K)$ colour steepens, such that galaxies
at the same stellar mass become redder. This effect is
dominated by the change in rest-frame wavelength
sampled by the $R$ and $K$ filters. The fact that the higher
mass galaxies become redder, while the lower mass galaxies
tend to remain blue, is a sign that the spectral energy 
distributions of the lower mass galaxies are bluer than those
for higher mass galaxies.  This pattern evolves however,
and at $z > 1$, galaxies at every mass bin become redder
with time.    The upper envelope in the colour-stellar mass
relation (Figure~8) is due to incompleteness, and is not a real limit.

On Figure~7 and Figure~8 we plot hydrodynamic simulation results
from Nagamine et al. (2005) using different dust extinctions.  We
over-plot the E(B-V) = 0, 0.4 and 1 models on these figures as contours. 
First, the fact
that there is not a larger scatter in the log M$_{*}$ vs. $K$ relation
(Figure~7) is an indication that these galaxies are not dominated by
dust extinction. Very few galaxies overlap with the E(B-V) = 1 model,
and most galaxies are better matched with the E(B-V) = 0 model. 
This can furthermore be seen in Figure~8 where
only the lowest extinction model, with E(B-V) = 0, generally matches
the location of galaxies in the $(R-K)$ vs. M$_{*}$ diagram which are not
EROs.  The (E-V) = 0.4 model does a good job of tracing the EROs, but
these systems could also be composed of old stellar populations.  We revisit
the issue of dust extinction in these galaxies in \S 5.1.

Finally, there are clearly two unique, and overlapping, samples identified in
this near-infrared selected sample.  The first are those
galaxies in Figure~7 and 8 which are very massive, with
masses M$_{*} > 10^{11}$ \solm. We discuss these objects, and
their evolution in Conselice et al.
(2007b).   We investigate in the next section the
properties of the extremely red objects (EROs), 
those with observed colours, $(R-K) > 5.3$.

\begin{figure*}
\hspace{-1cm}
%\vspace{5.5cm}
\vbox to 220mm{
\includegraphics[angle=0, width=190mm]{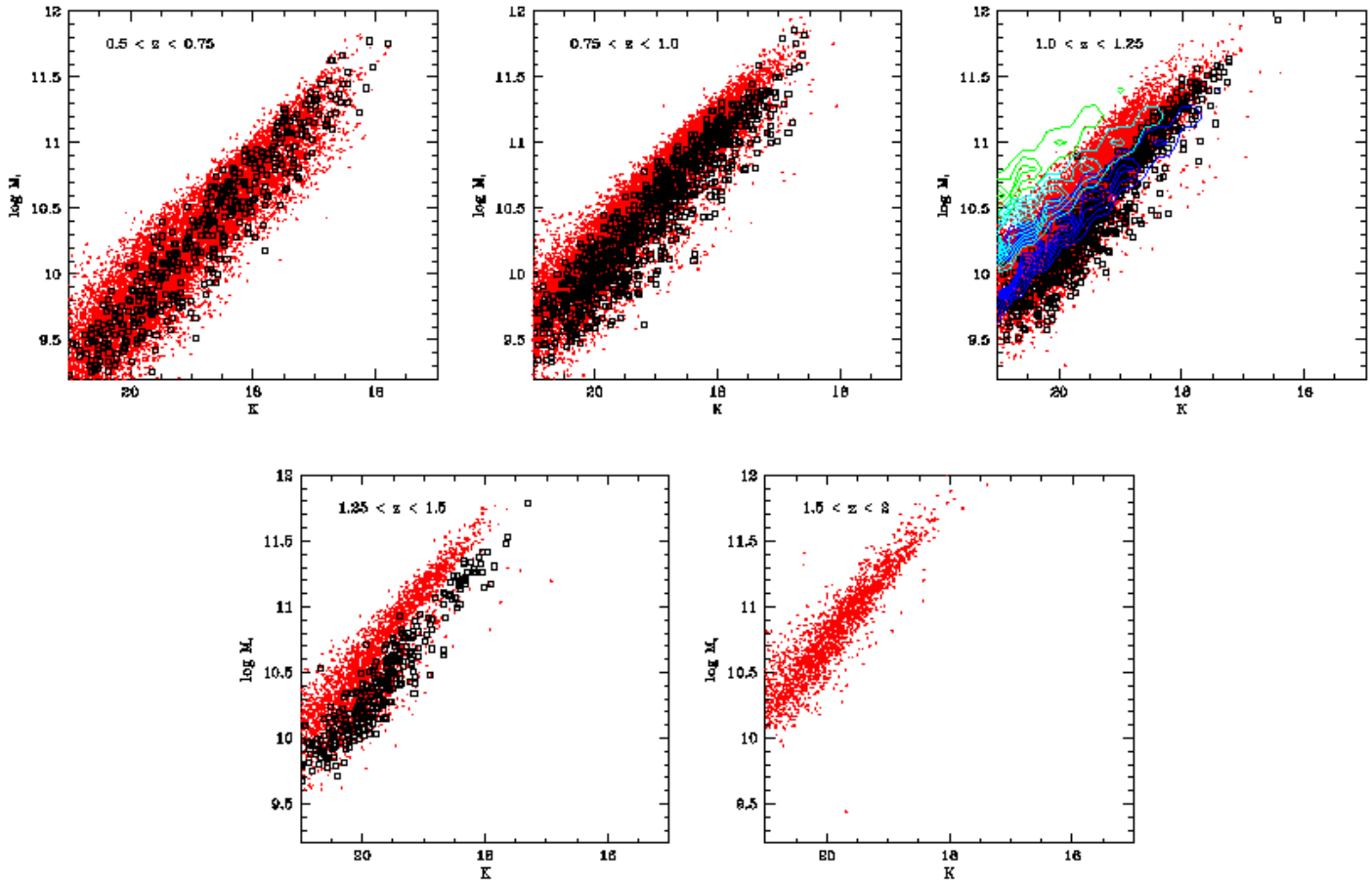}
%\vspace{-0cm}
 \caption{The stellar mass vs. K-band magnitude relation for
our sample of galaxies out to $z \sim 2$.  These figures are divided up into
different redshifts, increasing from left to right and top
to bottom.  Plotted on these figures are both systems
which have spectroscopically measured redshifts (open
black boxes) and those which have photometric redshifts (the
red dots).   As can be
seen, there is generally a strong relationship between
stellar mass and apparent K-band magnitude, with
a small scatter. Note that generally galaxies with
spectroscopically measured redshifts are those which are
brighter for a given stellar mass. These same systems
are furthermore on the blue edge of the stellar mass-colour
relationship (Figure~8). This shows that the DEEP2 spectroscopy
is selecting primarily the bluer and brighter galaxies at a given stellar
mass.  We also plot in the $1.0 < z < 1.25$ bin models for how
these quantities relate, from SPH simulations from Nagamine et al. (2005).  
The blue, cyan and green contours (going from low mass to high mass
at a given K) show the location of model galaxies 
with E(B-V) = 0, 0.4 and 1, respectively.   }
}  \label{sample-figure}
\vspace{-6cm}
\end{figure*} 

\begin{figure*}
\hspace{-1cm}
%\vspace{5.5cm}
\vbox to 220mm{
\includegraphics[angle=0, width=190mm]{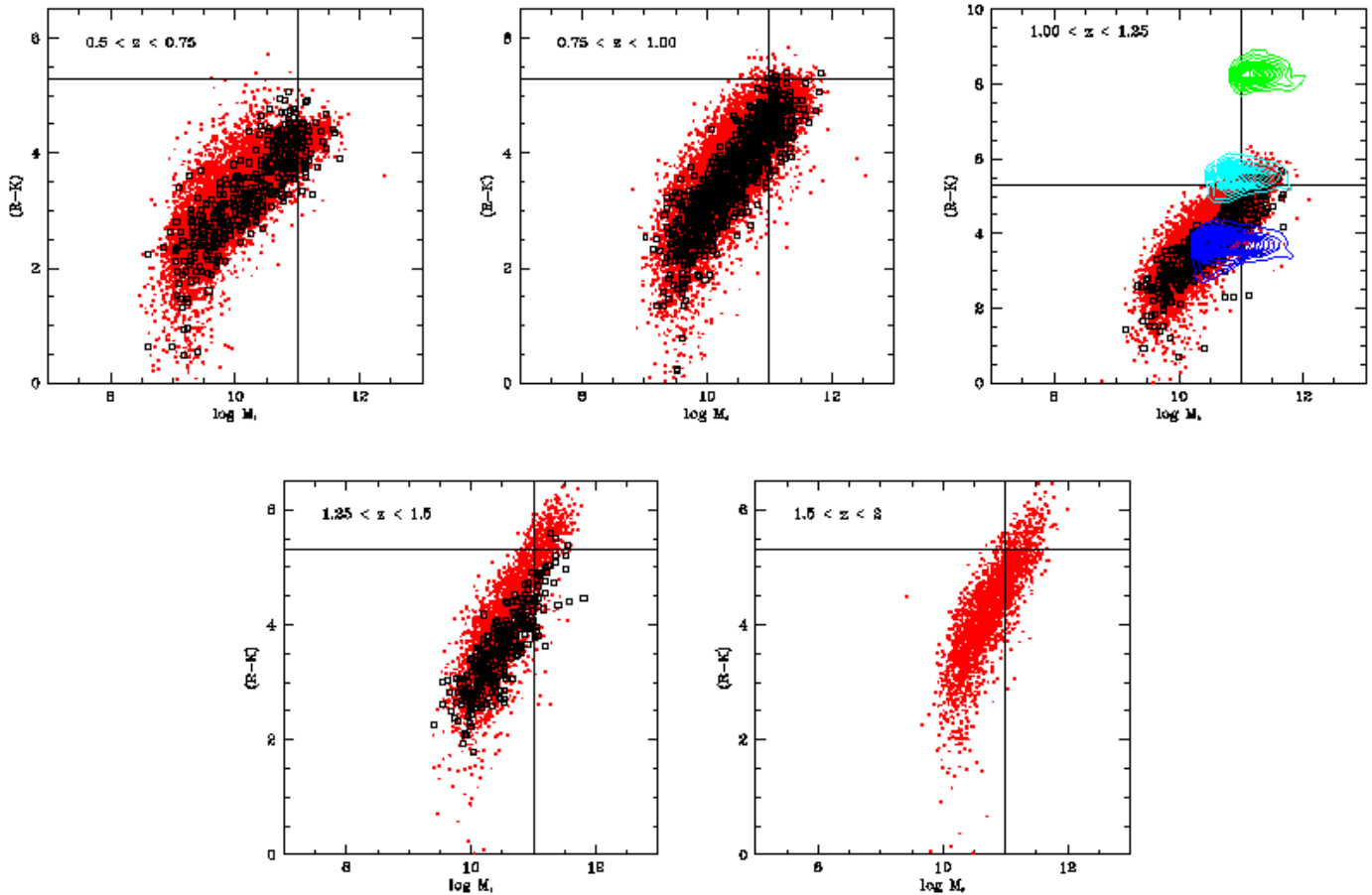}
%\vspace{-0cm}
 \caption{The colour-stellar mass relation for
our sample to $z \sim 0$.   Similar to
Figure~7, these galaxies have been divided up into
different redshift bins. The horizontal and vertical lines show
the limits for selecting unique galaxy populations probed
in the K-band. Galaxies to the right of the vertical line
are massive galaxies studied in
Conselice et al. (2007b), while galaxies which are above
the horizontal line, defined by $(R-K)>5.3$ are the EROs
studied in this paper.  Note that the final redshift panel
with $1.5 < z < 2.0$ consists solely of photometric redshifts.
Also, similar to Figure~7 we plot in the $1.0 < z < 1.25$ figure models 
results from Nagamine et al. (2005), although we utilise the model results
within the same limits as our data, $K < 20$.   As in Figure~7, the blue, 
cyan and green contours show the location of model galaxies with 
E(B-V) = 0, 0.4 and 1 (going from bluer to redder colours at a given
stellar mass), respectively.   }
} \label{sample-figure}
\vspace{-6cm}
\end{figure*}

\section{Properties of Extremely Red Objects}

\subsection{Sample Selection}

The ERO sample we construct is defined through a traditional colour
cut to locate the reddest galaxies selected in near infrared/optical
surveys with $(R-K) > 5.3$.   Galaxies selected in this way are
often considered  the progenitors of today's most massive galaxies, as
seen at roughly $z \sim 1-2$.  Objects with these extremely red 
colours have remained a population
of interest since their initial discovery (Elson, Rieke \& Rieke 1988).
Initially thought to be ultra-high redshift galaxies with $z > 6$, it is
now largely believed that EROs are a mix of galaxy types at $z > 1$
(e.g., Daddi et al. 2000; McCarthy et al. 2001; Firth et al. 2002;
Smail et al. 2002; Roche et al. 2003; 
Cimatti et al. 2002; Yan \& Thompson 2003; Cimatti et al. 2003; Moustakas et 
al. 2004; Daddi et al. 2004; Wilson et al. 2007). Generally, it has also 
been thought that EROs
should trace some aspect of the massive galaxy population (Daddi
et al. 2000); an idea we can test further with our data.
Furthermore, because EROs are an easily defined and observationally
based population, there has been considerable observation and
theoretical work done towards understanding these objects.  Naturally, it is
more desirable to work with mass selected sample (see Conselice
et al. 2007b for this approach), but these samples rely on accurate
redshifts and stellar mass measurements, while the EROs are simply
observationally defined through a colour.

The idea that EROs are red due to either an evolved galaxy
population, or a dusty starburst is perhaps no longer the dominate way to
think about these systems (e.g., Moustakas et al. 2004). However, there are
properties of EROs that are still not understood, nor
even constrained.  For example, it is not clear
why some EROs can have apparently normal galaxy morphologies, while
others appear to be merging or peculiar galaxies (e.g., Yan \&
Thompson 2003; Moustakas et al. 2004). 

The questions we address include: what are the stellar mass, morphological
and redshift distributions for these objects?  In our analysis
we study the properties of these traditionally colour 
selected EROs to determine these basic properties.

Our sample of EROs however is not defined simply by a $(R-K) > 5.3$ cut on our
entire $R-$band and $K-$band catalogs. Due to the depth of both
filters, we have to limit how deep we search for EROs to avoid false
positives.  As discussed in \S 2.4, we are 100\% complete in our entire survey
down to $K = 20$.  The $R-$band depth is however not well
matched to the $K$-depth for finding EROs, and has a $>$ 5 $\sigma$ detection
limit of $R = 25.1$ (although 50\% completeness).  We therefore only select 
EROs which we are certain 
to within $> 5$ $\sigma$ have a colour $(R-K) > 5.3$.  
This limits our analysis of EROs down to $K = 19.7$.

We divide our ERO sample into three types, depending on the
origin of the redshift for each. The first type are those EROs with a
high quality DEEP2 redshift, of which there are a total of 62
within our survey.  The second type of ERO are those with
$R < 24.1$ which contain an ANNz photometric redshift (\S 2.2).  There are
343 of these EROs.  The third type are the EROs with
magnitudes between $24.1 < R < 25.1$ which all have `full'
photometric redshifts (\S 2.2).
There are 1122 of these objects.  The entire  $(R-K) > 5.3$ sample therefore
consists of 1527 EROs with some type of redshift.   We examine
the properties of $(I-K) > 4$ selected EROs in \S 4.6.

\subsection{ERO Number Counts}

As with the number counts of the K-selected objects, we
are interested in comparing the number counts of our ERO
sample with measurements from previous work. In Figure~9 we plot 
the number counts
for our ERO sample, as a function of K-magnitude. As can
been seen, we are slightly under-dense at nearly all magnitudes
compared to the UKIDSS UDS survey, but find similar results as
Daddi et al. (2000).  The differences between the
counts in our survey and the UDS is likely due to several
issues, including the slightly
different filter sets used, and the correction for galactic
extinction.  

The UDS survey uses Subaru Deep
Field imaging utilising the Cousins $R_{\rm C}$ filter, while
our R-band imaging is from the CFHT and utilises a Mould R filter,
which has different characteristics.
Another issue is that these previous surveys have not corrected
for Galactic extinction, while we have. This can result in
slight differences in the number counts.
Another feature seen in previous surveys, which we also see,
is a turn-over in the slope of the counts at about K = 18.5,
towards a shallower slope at fainter magnitudes.

%ero.sm, numde
\begin{figure}
%\vspace{5.5cm}
 \vbox to 80mm{
\includegraphics[angle=0, width=84mm]{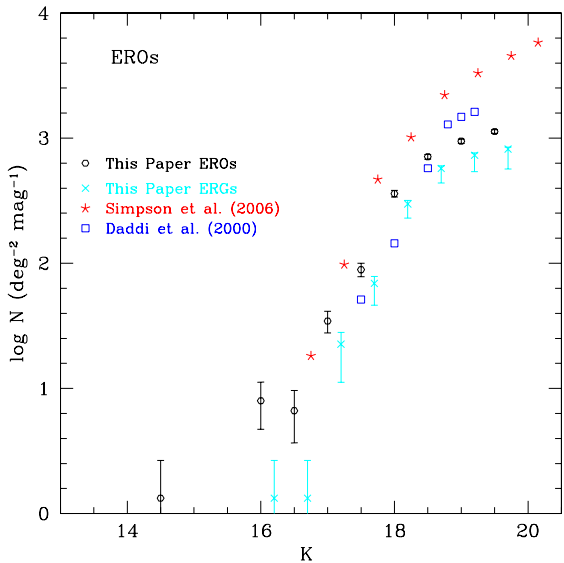}
%\vspace{-0cm}
 \caption{The number counts for EROs in our fields
plotted along with the counts from previous surveys
by Simpson et al. (2006) and Daddi et al. (2000).  We
also plot the number counts for the extremely red
galaxies (ERGs) - the ERO counts with stars removed.}

} \label{sample-figure}
\vspace{2cm}
\end{figure}

\subsection{Redshift Distributions and Number Densities of EROs}

One of principle quantities needed to understand the properties of
EROs is their redshift distribution and number densities.  With our 
three different types of redshifts (\S 2.2) measured for our EROs, we can 
construct the redshift distribution and number density evolution
for EROs down to a magnitude limit of $K = 19.7$.  Figure~10a shows the 
redshift distribution for our  $(R-K) > 5.3$ selected EROs.
We have plotted this distribution with three different histograms:
for the spectroscopic redshift sample (red diagonal hatch), a 
photometric redshift sample for galaxies at $R_{\rm AB} < 24.1$ (black),
and a photometric redshift sample
at $R_{\rm AB} > 24.1$ (blue horizontal dashed).   

It is clear that galaxies selected with the ERO criteria are at higher 
redshifts ($z > 1$), with very few galaxies meeting this criteria at $z < 0.8$,
and all of those that do are photometrically derived redshifts.   A similar
pattern can be seen in Figure~6, which plots the colour-redshift
distribution for our sample selected by $R_{\rm AB} < 24$,
and $R_{\rm AB} > 24$.

The average redshift for our $K < 19.7$ ERO sample is $<z> =$ 1.28$\pm$0.23.
This is at a higher redshift and has a smaller dispersion than the
average redshift for all galaxies with $K < 19.7$, $<z> = 0.84\pm0.31$. 
The above arguments show that the traditional $(R-K) > 5.3$ ERO cut 
reliably locates galaxies at $z > 1$, on average. There is also a fairly long
tail of sources up to $z \sim 2$.

\setcounter{table}{3}
\begin{table}
 \caption{Extremely Red Object Number Densities for Systems at $K_{\rm vega} < 19.7$.}
 \label{tab1}
 \begin{tabular}{@{}ccc}
  \hline
ERO Selection & Redshift & log ($\phi$) (h$^{3}_{70}$ Mpc$^{-3}$ dex$^{-1}$) \\
\hline
 $(R-K) > 5.3$   &  0.5 &  -6.00$^{+0.31}_{-0.33}$ \\
   &      0.7    &   -5.42$^{+0.16}_{-0.27}$ \\
   &      0.9    &   -4.64$^{+0.10}_{-0.13}$ \\
   &      1.1    &   -3.90$^{+0.09}_{-0.11}$ \\
   &      1.3    &    -4.10$^{+0.09}_{-0.11}$ \\
   &      1.5     &   -4.24$^{+0.10}_{-0.11}$ \\
   &      1.7    &    -4.70$^{+0.14}_{-0.12}$ \\
   &      1.9     &   -5.13$^{+0.13}_{-0.15}$ \\
\\
$(I-K) > 4$ &  0.5    &  -5.10$^{+0.15}_{-0.23}$ \\
    &     0.7    &  -5.19$^{+0.14}_{-0.21}$ \\
    &     0.9    &  -4.82$^{+0.11}_{-0.14}$ \\
     &    1.1    &  -4.16$^{+0.09}_{-0.11}$ \\
     &    1.3    &  -4.12$^{+0.09}_{-0.11}$ \\
   &      1.5    &  -4.11$^{+0.10}_{-0.11}$ \\
   &      1.7    &  -4.32$^{+0.13}_{-0.11}$ \\
   &      1.9    &  -4.47$^{+0.11}_{-0.11}$ \\
\hline
 \end{tabular}
\end{table}

For the most part it appears that EROs  at $K < 19.7$ are selecting high
redshift galaxies at $z \sim 1.3$. However, we are missing a few galaxies 
from our ERO sample at $K < 19.7$ which do not have
a redshift due to non-detections in the optical bands. These galaxies could
be at very high redshift, and will be discussed in a future paper.

Figure~11 plots the number density evolution for our EROs at $K < 19.7$ 
as a function of redshift, with tabulated values shown in Table~4.
As can be seen, in agreement with Figure~10, there
is a drop in the number density of EROs at $z < 1$.  The number density
peak for EROs is also clearly found between $z = 1 - 1.5$.  We can compare
this figure to previous measurements and models (e.g., Nagamine et al.
2005).   Previously Moustakas et al. (2004) and Cimatti
et al. (2002b) measured ERO number densities within various K-limits, 
but within $(R-K) > 5$, as opposed to our $(R-K) > 5.3$.   However, when we
compare our results to these papers, we find very similar
results.  Down to $K_{\rm vega} < 20.12$, Moustakas et al.
find a number density of EROs at $z = 1$ of log($\phi$(Mpc$^{-3}$)) = 
$-3.19$, whereas we find $-3.39$ in roughly good agreement. Similarly, 
Cimatti et al. (2002b) find down to $K_{\rm vega} < 19.2$ a density
of log($\phi$(Mpc$^{-3}$)) = $-3.67$ at $z = 1$, while we
find $-3.60$, again in good agreement.

When we compare our results to simulation results from
Nagamine et al. (2005) we find
again roughly good agreement, although the Lagrangian SPH results
find a slightly higher number density.  At $z = 1$ these SPH
simulations find log($\phi$(Mpc$^{-3}$)) = $-2.96$, while the 
total variation diminishing (TVD) simulations give a slightly higher
result.  This density is a factor of 2.6 higher than the number density
which we observe.  These density are however the result of assuming
a dust extinction of E(B-V) = 0.4, which might be too high in
light of the results shown in Figures~7 and 8.    Lower values of
E(B-V) will make galaxies less red, and will produce a lower
number density of EROs.

\subsection{Stellar Masses of EROs}

As shown in Figure~10b, our EROs generally have high stellar
masses, and thus a fraction of the  most massive galaxies at 
$z > 0.8$ must be EROs.  This is an important result, as it has
been surmised from other criteria, such as clustering, that the
EROs are contained within massive halos (e.g., Daddi et al. 2000; 
Roche, Dunlop \& Almaini 2003). 

We have constructed  a complete sample of
M$_{*} > 10^{11}$ \solm galaxies up to $z \sim 1.4$ in our fields (Conselice
et al. 2007b), from which we can directly test the idea
that EROs are massive galaxies.  Although Figure~10b demonstrates
that our EROs at $K < 19.7$ are selecting massive galaxies, this is 
likely due to the fact that our EROs are relatively bright, and we cannot 
constrain the masses or
redshifts of fainter EROs, which must be either lower mass
galaxies at the same redshifts as these, or higher redshift
massive systems.  There is also little difference in the
distributions of stellar masses for our ERO sample at
different redshifts.  The peak mass is around 
2$\times 10^{11}$ \solm at all redshift selections.

  We find that most of 
our sample of $K < 19.7$ EROs tend to have masses M$_{*} > 10^{11}$ \solm,
which in the nearby universe are nearly all early-types (Conselice
2006a).    This is a strong indication that $K < 19.7$ EROs, regardless of 
their
morphology or stellar population characteristics, are nearly certain to
evolve into passive massive early-types in the nearby universe.

\subsubsection{Are Massive Galaxies at $z > 1$ EROs?}

\begin{figure*}
%\vspace{5.5cm}
 \vbox to 150mm{
\includegraphics[angle=0, width=154mm]{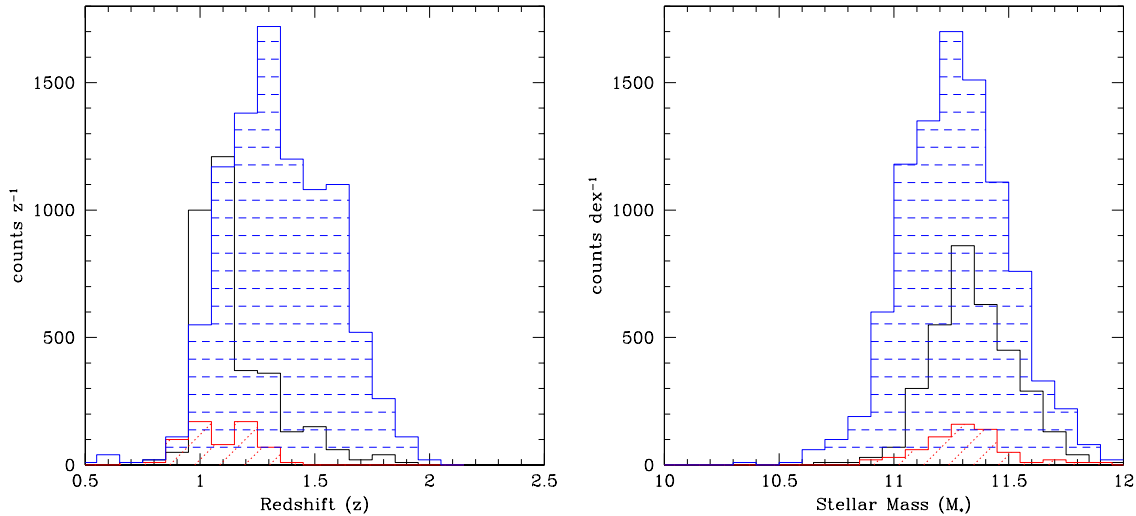}
 \caption{The redshift and stellar mass histograms for
our sample of EROs. Shown are three histograms created after dividing
the sample three different ways, depending on the
origin of the redshift. The dotted
diagonal hatched 
red histogram shows the redshift and stellar mass
distributions for
the spectroscopically confirmed EROs, while the
non-hatched black histogram shows the distributions for
EROs with $(R-K) > 5.3$, $R_{\rm AB} < 24.1$, and whose redshifts
are photometric.  The horizontal blue dashed hatched
histogram shows the distributions for EROs with
$R_{\rm AB} > 24.1$ with measured photometric redshifts. As can
be seen at a $K < 19.7$ limit, the ERO selection generally
locates massive galaxies at $z > 1$.  }
} \label{sample-figure}
\vspace{-2cm}
\end{figure*} 

While EROs are massive galaxies, the opposite of this is not necessarily
true, as there are 
massive galaxies with M$_{*} >$ \mass, that are not EROs, some with
very blue colours (Conselice et al. 2007b). Figure~12 plots
the fraction of galaxies within the mass ranges M$_{*} >$ \hmass and 
\mass $<$~M$_{*}<$ \hmass which are EROs between $z \sim 0 - 2$.
Figure~11 plots the number density evolution for EROs selected
in two ways and compares to galaxies selected with stellar 
masses M$_{*} >$ \mass.
There are several interesting features in these figures. 
The first is that while massive galaxies exist throughout
this redshift range, EROs only populate massive galaxies
at $z > 1$. Another interesting feature is that an ERO selection
at $K < 19.7$ will include a large fraction of the most massive
galaxies with M$_{*} >$ \hmass, at $1 < z < 2$.   On average, between
$1.0 < z < 1.4$, our ERO selection will find 36\% of all M$_{*} >$ \hmass
galaxies.  This increases to 75\% within the redshift range
$1.2 < z < 1.8$.  

However, the ERO colour limit does not do a good job
in selecting massive galaxies with \mass $<$~M$_{*} <$ \hmass.
In this mass range at $1.0 < z < 1.4$ the ERO selection finds
only 35\% of these systems. Similar to the M$_{*} >$ \hmass
mass range, we find a higher fraction of  \mass $<$~M$_{*}<$ \hmass
galaxies which are EROs at 44\%. However, with a $K < 19.7$ limit,
we are obtaining a similar number density of EROs per co-moving volume
as there are massive galaxies (Figure~11). The number densities
of EROs however declines rapidly at $z < 1$.

While it appears that the traditional $(R-K) > 5.3$ limit
will find the most massive galaxies at $z > 1$, this colour cut
does not give a purely ultra-high mass sample, nor does it include
all of the massive galaxies at these redshifts.  At $1.2 < z < 1.8$ about
25\% of galaxies with M$_{*} >$ \hmass and 66\% of systems with
\mass $<$~M$_{*} <$ \hmass are not EROs.  This is consistent with
our finding in Conselice et al. (2007b) that $\sim 40$\% of
massive galaxies at $z > 1$ are undergoing star formation and have
blue $(U-B)_{0}$ colours.

\subsection{Structural Features}

We utilise visual estimates of Hubble types and 
the non-parametric CAS system to characterise the morphologies and
structures of an ERO sample selected with $(R-K) > 5$. While there
are slightly more systems at $(R-K) > 5.3$ than at
$5 < (R-K) < 5.3$, we reduce our limit to aquire more
systems for analysis and to better compare with previous work.   
A similar analysis is
the focus of Conselice et al. (2007a,b), in which we examined
the morphological properties of the most massive galaxies at $z > 1.5$,
as well as the Distant Red Galaxies (DRG), with $(J-K) > 2.3$, which
are also proposed to be the progenitors of today's massive galaxies.

%ero.numd.sm, mevol in /anal8/
\begin{figure}
%\vspace{5.5cm}
 \vbox to 80mm{
\includegraphics[angle=0, width=84mm]{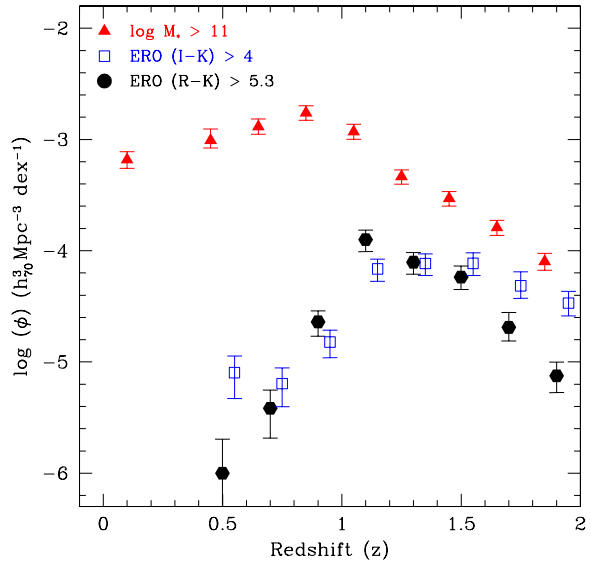}
%\vspace{-0cm}
 \caption{The co-moving number densities of EROs selected by 
the $(R-K)_{\rm vega} > 5.3$ and the $(I-K)_{\rm vega} > 4$ criteria as a 
function of
redshift. Also plotted for reference is the number density evolution
for galaxies with stellar masses M$_{*} >$ \mass (see Conselice
et al. 2007b).}

} \label{sample-figure}
\vspace{4cm}
\end{figure}  

The CAS (concentration,
asymmetry, clumpiness) parameters are a non-parametric method for
measuring the structures of galaxies on CCD images (e.g., Conselice 2007;
Conselice et al. 
2000a,b; Bershady et al. 2000; Conselice et al. 2002; Conselice 2003).  The
basic idea is that low redshift, nearby galaxies, have light distributions
that reveal their past and present formation modes (Conselice 2003). 
Furthermore, well-known galaxy types in the nearby universe fall in 
well defined regions of the CAS parameter space (Conselice
2003).   We apply the CAS system to our EROs to determine their
structural features.  There are two caveats to using the ACS imaging
on these galaxies.  The first is that there are redshift
effects which will change the measured parameters, such that the asymmetry
and clumpiness indices will decrease (Conselice et al. 2000a; Conselice
2003), and the concentration index will be less reliable (Conselice
2003).  There is also the issue that for systems at $z > 1.3$ we
are viewing these galaxies in their rest-frame ultraviolet images, which
means that there are complications when comparing their measured structures
with the rest-frame optical calibration indices for the nearby galaxies.
Our main purpose in using the CAS system is to identify relaxed massive
ellipticals as well as any galaxies that are still involved in a recent
major merger and are presumably dusty.

The following structural and morphological analysis is based on the ACS
imaging of the EGS field described in \S 2.1.  The imaging we use
covers 0.2 deg$^{2}$ in the F814W (I) band, giving us coverage
for $\sim 15$\% of our ERO sample. 

%ero.masse.sm, mevol
\begin{figure}
%\vspace{5.5cm}
 \vbox to 80mm{
\includegraphics[angle=0, width=84mm]{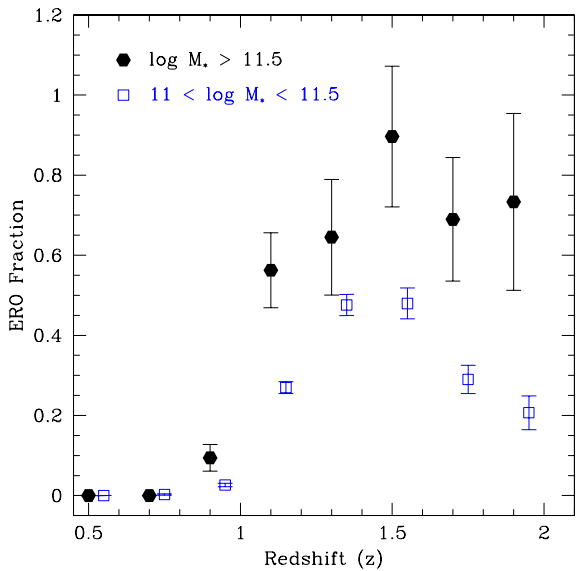}
%\vspace{-0cm}
 \caption{Diagram showing the fraction of galaxies with 
extreme masses M$_{*} >$ \hmass (solid circles)  and
\mass $<$~M$_{*}<$ \hmass (open boxes) which are also $(R-K) > 5.3$ 
selected EROs. As can be seen, the ERO selection successfully finds
galaxies at $z > 1$, yet does not locate all of these systems. A sample
of EROs at $K < 19.7$ will contain nearly all of the log M$_{*} > 11.5$
systems at $z \sim 1.5$, but is less successful at finding the lower
mass, log M$_{*} > 11$ systems.}

} \label{sample-figure}
\vspace{4cm}
\end{figure}

\subsubsection{Eye-Ball/Classical Morphologies}

We study the structures and morphologies of our sample using two different
methods.   The first is through simple visual estimate of morphologies based
on the appearance of our ERO sample in CCD imaging. The outline of this
process is given in Conselice et al. (2005a) and is also described
in the companion paper (Conselice et al. 2007b).  Our total sample of 
objects gives 436 unique EROs for which there is ACS imaging.  Each of these 
galaxies were 
placed into one of six categories: compact, elliptical, lenticular (S0),
early-type disk, 
late-type disk, edge-on disk, merger/peculiar, and 
unknown/too-faint. These classifications are very simple, and are only based 
on appearance. No other information, such as colour or redshift, was used to
determine these types. An outline of these types is provided below, with
the number in each class listed at the end of each description.

\noindent 1. Ellipticals: A centrally concentrated galaxy with no evidence 
for a lower surface brightness, outer structure (152 systems). An additional
58 galaxies were classified as peculiar ellipticals, which appear
similar to ellipticals, but have an unusual light distribution, or
bulk asymmetry (see Conselice et al. 2007b).

\noindent 2. Lenticular (S0): A galaxy was classified as an S0 if it
appears as an elliptical but contained a disk-like outer structure with 
no evidence for spiral arms, or clumpy star forming knots or other
asymmetries. (13 systems)

\noindent 3. Compact - A galaxy was classified as compact if its structure
was resolved, and is very similar to the elliptical classification in that
these systems must be very smooth and symmetric. A compact galaxy differs
from an elliptical in that it contain no obvious features, such as an extended
light distribution or envelope.  (24 systems)
     
\noindent 4. Early-type disks: If a galaxy contained a central 
concentration with some evidence for lower surface brightness outer 
light, it was classified as an early-type disk. (3 systems)
 
\noindent 5. Late-type disks: Late-type disks are galaxies that appear to 
have more outer low surface brightness light than inner concentrated light. 
(1 systems)

\noindent 6. Edge-on disks: disk systems seen edge-on and whose face-on
morphology cannot be determined but is presumably S0 or disk. (17 systems)
    
\noindent 8. Peculiars/irregulars: Peculiars are systems that appear to be 
disturbed or peculiar looking, including elongated/tailed sources. These 
galaxies are 
possibly in some phase of a merger (Conselice et al. 2003a,b) and are common at 
high redshifts. (148 systems)
   
\noindent 9. Unknown/too-faint: If a galaxy was too faint for any reliable 
classification it was placed in this category. Often these galaxies appear 
as smudges without any structure. These could be disks or ellipticals, but 
their extreme faintness precludes a reliable classification. (20 systems)

\subsubsection{Morphological Distributions}

The morphological distribution of the  EROs can help us
address the question of the origin of these extremely red galaxies. In the
past, this technique has been used to determine the fraction of
EROs which are early-type, disk or peculiar.  Previous studies 
on this topic include Yan \& Soifer (2003) who study 115 EROs, 
and Moustakas et al. (2004) who studied 275 EROs in the GOODS
fields. Our total population of EROs for which we have
morphologies is 436.  These earlier studies have found a mix of types,
with generally half of the EROs early-types, and the other
half appearing as star forming systems in the form of disks or
peculiars/irregulars. 

In summary, we find that 57$\pm$3\% of our EROs 
are early-type systems. This includes 58 systems, or 13\% of
the total, which are disturbed ellipticals. In
classifications carried out in previous work some of these 
systems would be classified as peculiars.  The bulk of the rest of
the population consists of bonafide peculiars, which make up
34$\pm$3\% of the ERO population. Only four EROs
were found to be face on disk galaxies, while
17 systems (4\%) of the ERO sample are made up of
edge-on disk galaxies.  Presumably these galaxies
are red for different reasons - either evolved galaxy
populations, or dusty star formation, or from dust
absorption produced through orientation in the case
of edge-on disks.

Previous work has been somewhat inconsistent on the
morphological break-down between peculiars and
early-type galaxies (e.g.,  Yan \& Soifer 2003; Moustakas et al. 2004). 
From our study, it is clear
that much of this difference can be accounted for
by the peculiar ellipticals. These systems appear
in their large-scale morphology to be early-type,
but have unusual features, such as offset nuclei
that make them appear peculiar.  The differences
between previous findings can largely be accounted for
by whether these peculiar ellipticals were included
in the early-type or peculiar class.

\begin{figure}
\includegraphics[width=84mm]{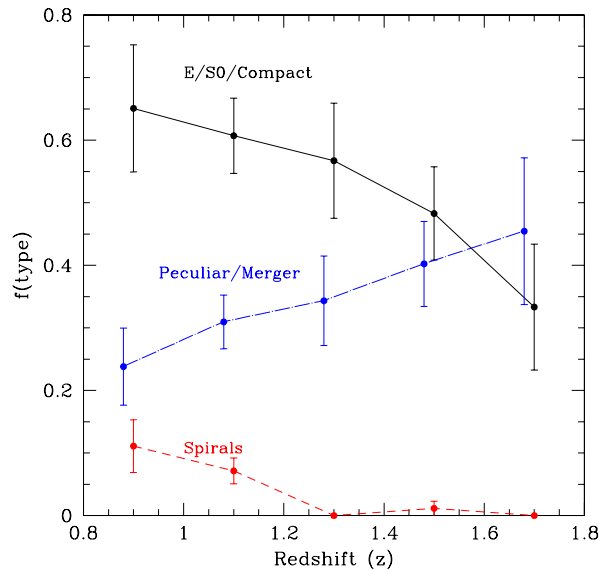}
  %% to include a figure, or
% \vspace{3.5cm}
  %% to leave a blank space
 \caption{Our visual estimates of ERO morphological type
as a function of redshift.  As labelled, the ellipticals, S0s and
compact morphological types are shown as a solid black line. The
systems classified as a peculiar/mergers are shown as the dotted
blue line and spirals are shown as the dashed red line.}
 \label{sample-figure}
\end{figure}

We find that the relative number
of peculiar and early-type EROs changes with redshifts
(Figure~13), such that at the lowest redshifts ($z \sim 0.7$)
the EROs are dominated by the E/S0/Compacts, with a type
fraction of $\sim 65$\%, but at $z > 1.4$
the peculiar systems are more prominent. The fraction of
EROs which are peculiars evolves from $\sim 45$\% at $z \sim 1.7$
to $\sim 20$\% at $z = 0.9$.  This mix between
early types and peculiars evolves with redshift, although
the exact reason for this evolution is not immediately clear. It
is possible that some of the peculiars at $z > 1.2$ only
appear so because we are sampling their morphologies below
the Balmer break that
would produce more irregular/peculiar looking galaxies
at these redshifts (Windhorst et al. 2002; Manger-Taylor et al.
2006; Conselice et al. 2007b).  However, we are nearly always
probing the rest-frame optical where the effects of the morphological
k-correction are minimised both qualitatively and quantitatively
(e.g., Conselice et al. 2005a; Conselice et al. 2007c in prep).  
We also find a higher
fraction of peculiar systems within the ERO sample, than
what is found for massive galaxies with M$_{*} >$ \mass
at similar redshifts (Conselice et al. 2007b).  

Interestingly, we find that 10-15\% of EROs at $z \sim 0.9$ are
spiral/disks.  For the most part however, it 
appears that most of the EROs are ellipticals, but we find that
peculiars make up roughly a third of the systems, 
with the relative contribution coming from galaxies at the
highest redshifts. We discuss the morphological break-down
of these systems in \S 5.2 in terms of models of galaxy
formation and evolution.

\subsubsection{CAS Structural Parameters}

Another way to understand the structures of these systems is
through their quantitative structural parameters as measured
through the CAS system. The CAS parameters have a well-defined range of 
values and are computed using simple techniques.  The concentration index is 
the logarithm of the ratio of the radius containing 80\% of the light in a 
galaxy to the radius which contains 20\% of the light (Bershady et al. 
2000).   The range in $C$ values is found from 2 to 5, with higher $C$ 
values for more concentrated galaxies, such as massive early types.  The 
asymmetry is  measured by rotating a galaxy's image by 180\deg and 
subtracting this rotated image from the original galaxy's image.  
The residuals of this subtraction are compared with the original galaxy's 
flux to obtain a ratio of asymmetric light.  The radii and centreing 
involved in this computation are well-defined and explained in 
Conselice et al. (2000a).  The asymmetry ranges from 0 to $\sim 1$ with 
merging galaxies typically found at $A > 0.35$.   The clumpiness is defined 
in a similar way to the asymmetry, except that the amount of light in high 
frequency `clumps' is compared to the galaxy's total light (Conselice 2003). 
The values for $S$ range from 0 to $> 2$, with most star forming galaxies
having values, $S > 0.3$.

We show in Figure~14 the CA and AS projection
of CAS space for EROs defined by $(R-K) > 5$.  As can be seen, the EROs, 
which are mostly early-types and peculiars, as defined by
eye (Figure~14), fall along a large portion of the range of possible values 
in CAS space. As expected, the irregulars/peculiars have higher 
asymmetries, lower concentrations, and higher clumpiness values than
the early types.  This is similar to, but not exactly the same,
as the structural parameter distribution for the most massive galaxies
with M$_{*} >$ \mass at $z < 1.4$.  There is a larger fraction of
peculiars within the ERO sample, and the redshift evolution with
types is not identical between EROs and massive galaxies 
(cf. Conselice et al. 2007b).

As with the massive galaxies found at high redshifts, the ERO
visually determined types are slightly higher in asymmetry than
their $z \sim 0$ counterparts (Conselice 2003).
Figure~14 shows how the EROs classified as early-types have
slightly larger asymmetries than their visual morphology would
suggest. The distorted early-types have even higher asymmetries
on average.  Most systems also deviate from the asymmetry-clumpiness 
relation (Figure~14), showing that the production of
these asymmetries is more likely due to dynamical effects,
rather than star formation (Conselice 2003). Interestingly,
we find that many of the peculiar EROs do not have a high
clumpiness index, which is opposite to what we found
for high asymmetry systems within the massive galaxies
sample (Conselice et al. 2007b).  The reason for this is that
these red galaxies are likely dusty, and therefore bright
star clusters are not seen within the ongoing star formation.
This furthermore shows that the EROs are likely more dominated by
galaxy merging than a pure mass selected sample.

\begin{figure*}
%\vspace{5.5cm}
 \vbox to 150mm{
\includegraphics[angle=0, width=154mm]{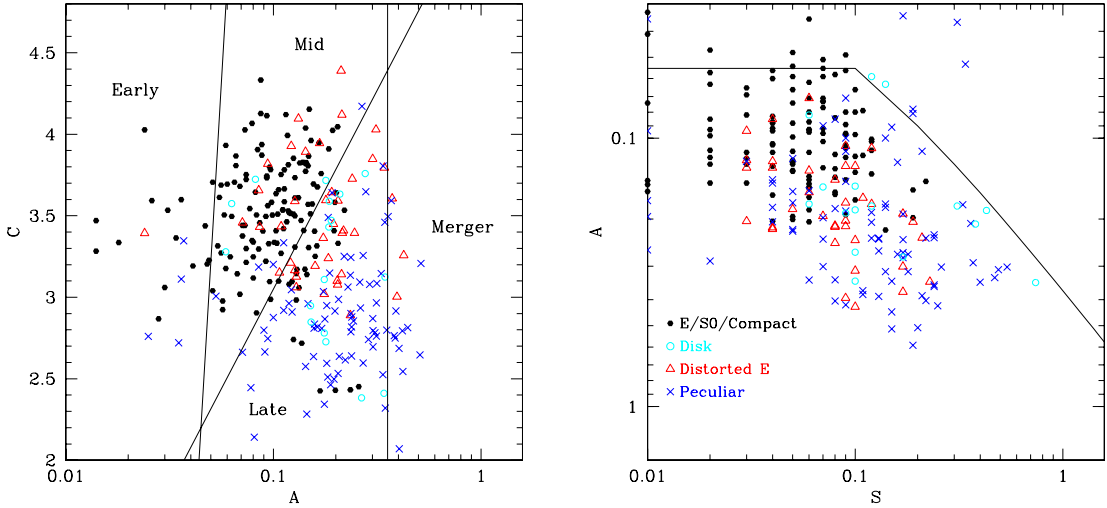}
%\vspace{-0cm}
 \caption{The CAS space for our ERO sample with $(R-K) > 5$.
Objects are labelled by their visual classification,
discussed in \S 4.5.1.   The solid black symbols represent the normal
(non-distorted) ellipticals, S0s and compacts.  The open red
triangles are the distorted ellipticals, while the blue crosses are
galaxies classified as peculiars/mergers, and the open cyan circles
are for the few disk galaxies found within the ERO sample.}
} \label{sample-figure}
\vspace{-6cm}
\end{figure*} 

We can understand the origin of these morphologies, and how
they relate to the origin of the EROs, by comparing the
CAS parameters of the EROs to their stellar masses.  Figure~15 shows
the comparison between ERO asymmetries and concentrations vs. their
stellar masses. The concentration-stellar mass diagram shows
a few interesting trends.  The most obvious feature is that
there appears to be a broad bimodality between EROs which are
peculiar, and those which are early-types. 

The peculiars and
early-types have similar masses, typically M$_{*} \sim 10^{11}$
\solm, yet they have different light concentrations.  The peculiars
typically have lower concentrations, $C = 2 - 3$, while the early-types
are generally at $C > 3$.  The early-types also tend to show
a correlation between concentration and stellar mass which is
not seen for the peculiars. The distorted ellipticals
fall in between these two populations suggesting an evolutionary
connection in the passive evolution of galaxy structure.  What 
we are potentially witnessing is the gradual transition from
peculiar EROs at high redshifts, to passive ellipticals at lower
redshifts, while on the way going through a distorted elliptical
stage. This is consistent with the idea that the $z = 1-2$ epoch
is where massive galaxies final reach their passive morphology
(Conselice et al. 2003; Conselice et al. 2005; Conselice 2006).

\begin{figure*}
%\vspace{5.5cm}
 \vbox to 150mm{
\includegraphics[angle=0, width=154mm]{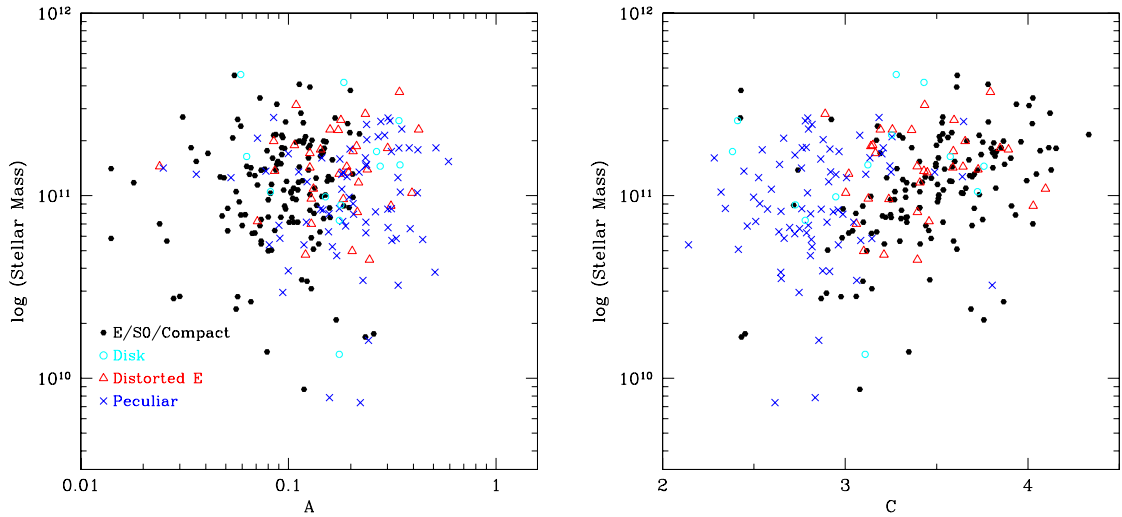}
 \caption{The relationship between asymmetries and light
concentrations of our ACS EROs with their stellar masses. The
symbols for each type are the same as in Figure~14.  There appears
to be a remarkable bimodal distribution in these parameter spaces
such that peculiars have low light concentration and higher asymmetries
than the elliptical-like objects, while having similar stellar masses.
The distorted ellipticals appear to fall in the gap between these
two populations, and are likely a transitional phase between the two.
This can be seen furthermore in Figure~13 where the fraction of
early-type EROs increases at the expense of the peculiars.}
} \label{sample-figure}
\vspace{-4cm}
\end{figure*}

There also appears to be a bimodality within the stellar mass/asymmetry
diagram (Figure~15).   The peculiars in general have higher asymmetries
than the early-types, but with the distorted ellipticals containing
asymmetries mid-way between the peculiars and the ellipticals, suggesting
again  that
the distorted ellipticals are a mid-way point in the evolution between
the peculiar EROs and the passive EROs.  As we do not see much mass
evolution in the upper edge of the ERO mass distribution, it is likely
that the peculiars are within their final merger phase, and  transform 
into early-types relatively quickly over $1 < z < 2$.  A similar pattern
can be seen for a high mass selected sample of galaxies at similar
redshifts (Conselice et al. 2007b).

\subsection{Other ERO Selection Criteria}

ERO selection through the $(R-K) > 5.3$ criteria is only one way
to find extremely red objects. Another popular method for finding EROs
is a selection with $(I-K) > 4$ (e.g., Moustakas et al. 1997).  
While both of these selection
methods are used to find EROs, it is not clear how these two  
methods compare, and whether they are finding the same galaxy 
population.  We investigate this issue briefly in this section.
 
\begin{figure}
%\vspace{5.5cm}
 \vbox to 80mm{
\includegraphics[angle=0, width=84mm]{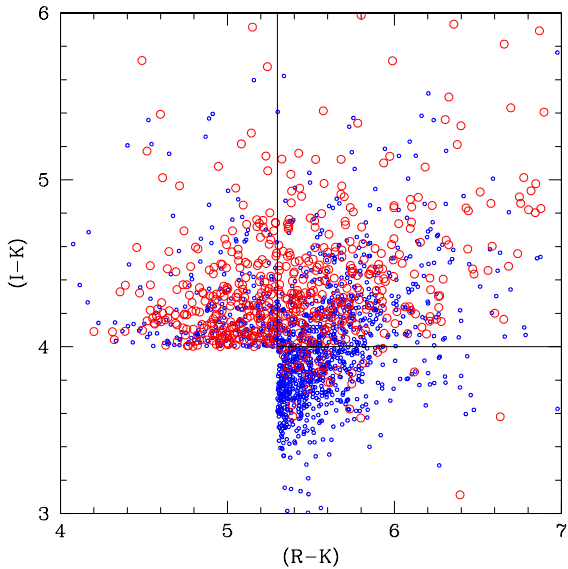}
 \caption{The $(I-K)$ vs. $(R-K)$ diagram for galaxies in our sample.
Only shown are those galaxies which are considered EROs either
through the $(R-K) > 5.3$, or the $(I-K) > 4$ criteria.  The points are
plotted in terms of their redshifts, with galaxies at $z > 1.5$ shown as 
open red circles and galaxies at $z < 1.5$ as blue dots.  
The $(I-K) > 4$ limit appears to find galaxies at higher redshifts, 
on average, than the $(R-K) > 5.3$ limit.} 
}\label{sample-figure}
\vspace{4cm}
\end{figure}

Figure~16 shows the relationship between $(I-K)$ and $(R-K)$ colours
for galaxies within our sample.  Those objects which have photometric
redshifts $z > 1.5$ are plotted as the red open symbols, while
those at $z < 1.5$ are plotted as the blue dots.  What is obvious from
this figure is that an ERO selection with $(I-K) > 4$ is more likely
to pick out galaxies at higher redshifts than the $(R-K) > 5.3$ limit.
This limit also appears to contain more contamination from lower
redshift galaxies than the $(R-K) > 5.3$ limit.  Overall, we find
an overlap of 767 EROs through both definitions down to $K = 19.7$,
this overlap constitutes 54\% of the $(R-K) > 5.3$ EROs and 60\% of the 
$(I-K) > 4$ EROs.

In Figure~17 we show the redshift distribution for the EROs selected
with $(I-K) > 4$, which can be directly compared with the redshift
distribution for $(R-K) > 5.3$ systems in Figure~10. As can be seen, 
the redshift distribution
for the $(I-K) > 4$ systems is skewed towards higher redshifts
than the $(R-K) > 5.3$ selection. We find that, on average, the
redshifts for the $(I-K) > 4$ EROs is $<$z$>$ = 1.43$\pm0.32$, compared
with $<$z$>$ = $1.28\pm0.23$ for EROs selected with the $(R-K) > 5.3$ limit.
There are also fewer systems at $z < 1.4$ within the
$(I-K) > 4$ selection than for the $(R-K) > 5.3$ limit, suggesting
that the redder bands are finding galaxies at slightly higher
redshifts.  However, this does not appear to be the case for the
$(J-K)>2.3$ `distant red galaxies' (DRGs), as discussed in
Conselice et al. (2007a) and Foucaud et al. (2007).  It appears
that these systems are picking up a significant fraction of
massive galaxies at $z \sim 1$ up to $K = 20$.

\begin{figure}
%\vspace{5.5cm}
 \vbox to 80mm{
\includegraphics[angle=0, width=84mm]{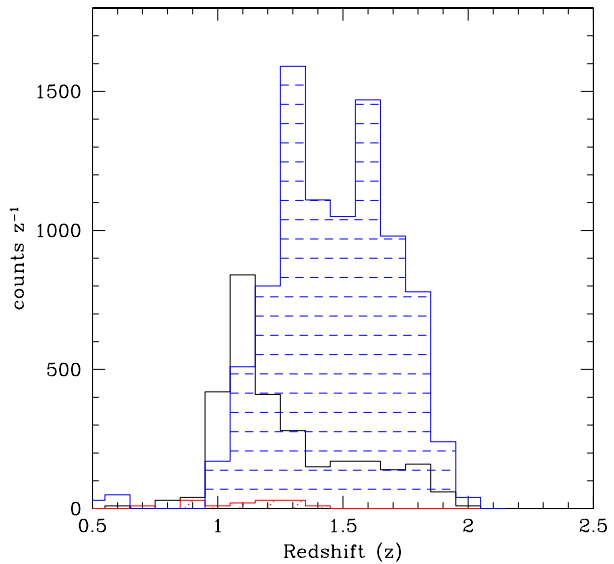}
 \caption{Similar to Figure~10, but for ERO systems
which have been selected with the $(I-K) > 4$ limit. As can
be seen by comparing this figure with Figure~10, we are
selecting, on average, higher redshift galaxies with these
redder bands.}
} \label{sample-figure}
\vspace{2cm}
\end{figure} 

\section{Discussion}

Here we discuss the results from this paper in the context
of galaxy formation models and scenarios. We include in this
discussion the redshift distribution of K-selected galaxies,
as well as the stellar mass and structural/morphological properties
of EROs to address how the stellar mass
assembly of galaxies is likely taking place.  By examining these
galaxies we are not relying on assumptions about stellar masses
to find the most massive and evolved galaxies at high redshift. In
this sense a K-band selected and colour selected population are
an alternative approach from stellar mass selection (Conselice
et al. 2007b) for understanding galaxy formation due to
the simplicity, and reproducibility, of their selection.

We first examine the redshift  distribution of $K < 20$ galaxies, and compare
this to models. We use colour information of our faint K-selected galaxies
to rule out all monolithic collapse formation scenarios for 
galaxies. We then examine the properties of the EROs themselves to further
argue that these systems are, due to their stellar masses, likely
an intrinsically homogeneous population, with the peculiar 
EROs evolving into the ellipticals at lower redshifts.

\subsection{K-band Redshift Distributions}

The number of K-band selected galaxies per redshift at a given K-magnitude
limit is an important test of when the stellar masses
of galaxies were put into place.  In general,
ideas for how massive galaxies form explicitly predict how much
stellar mass galaxies would have in the past. In
a rapid formation, such as with a monolithic collapse,
the stellar mass for the most
massive galaxies, which in a K-band limited sample will always
probe the massive systems, the number of galaxies within
a bright K-band limit, say $K < 20$ at high redshift should be much larger
than the number of sources seen in a hierarchical model, which
predicts that the stellar masses of galaxies grows with time.
Therefore the number of bright galaxies at higher
redshifts in a hierarchical model is less than  
that predicted in a monolithic collapse.  This test was
first performed by Kauffmann \& Charlot (1998) who concluded,
based on an early hierarchical model, that the predicted
counts exceed the observations, a result which has
remained despite improvements in data and models (e.g., Kitzbichler
\& White 2006).

\begin{figure}
%\vspace{5.5cm}
 \vbox to 90mm{
\includegraphics[angle=0, width=84mm]{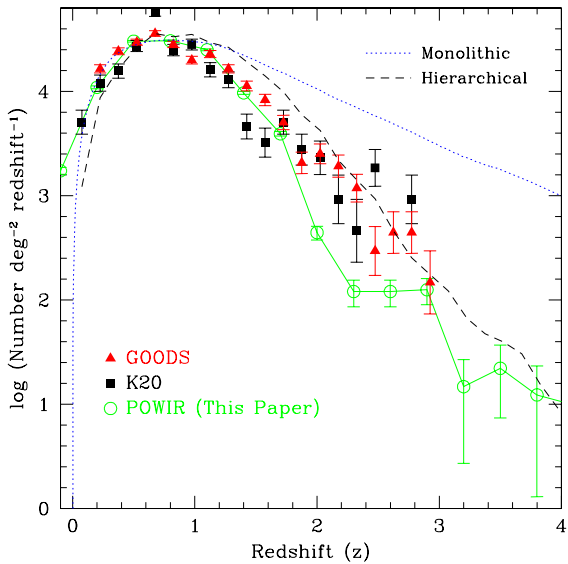}
 \caption{Redshift distribution of our sample
with a $K < 20$ cut.  Comparisons to previous redshift
distributions published in the K20 and GOODS surveys
are shown.  The two lines demonstrate predictions of fiducial
monolithic collapse (Kitzbichler \& White 2006), and hierarchical 
model predictions (Kitzbichler \& White 2007) for this evolution
within the same K-band limit.}
} \label{sample-figure}
\vspace{2cm}
\end{figure}

In Figure~18 we show the number of $K < 20$ galaxies
as a function of redshift for systems at $z < 4$. We further
plot on Figure~18 model predictions for how $N(z)$ changes in a standard pure
luminosity evolution monolithic collapse model from Kitzbichler \& White 
(2006), and within a standard hierarchical 
formation scenario from Kitzbichler \& White (2007).  We have used in this
figure our entire K-band distribution, including
galaxies for which we had to measure photometric redshifts without
an optical band (\S 2.2). We also plot on Figure~18 the $K < 20$ magnitude
distribution for galaxies seen in the GOODS and K20 surveys.  

While we 
generally agree with these previous results, we find a slightly lower
number of systems at higher redshifts compared to GOODS and K20. The
reason for this could simply be cosmic variance, as the GOODS and K20
samples at these redshifts also differ by a large amount. Otherwise, this
difference is likely created by errors within our photometric redshifts. 
However, it must be noted that the integrated
number of $K < 20$ galaxies is similar for our survey and GOODS and K20,
but is still much smaller than that predicted by the monolithic model.

As can be seen in Figure~18 there is clearly a large difference
between the observed distribution, and the predicted  monolithic collapse
distribution. We can rule out this basic monolithic collapse model, which
does not include dust, at $>$
10 $\sigma$, based on the comparisons to our $K < 20$ redshift
distribution.  It is possible to match with a monolithic collapse
model if extreme dust content is included in these galaxies, or if
there are `hidden' galaxies (Kitzbichler \& White 2006).  However,
as we argue below, there is no evidence that dusty galaxies dominate
our K-band selected sample.

 From Figure~18 it appears that a basic 
hierarchical model agrees better with the data (e.g.,
Somerville et al. 2004; Stanford et al. 2004; Kitzbichler \& White
2007).  There is perhaps a slight excess of galaxies in the hierarchical
formation model, which has been noted before (e.g., Kauffman \& Charlot 1998;
Somerville et al. 2004;  Kitzbichler \& White 2007).  The origin of
this difference is not clear. It perhaps simply implies that too much
mass is produced early in the hierarchical model, yet this would
seem to contradict the fact that the most massive galaxies with
M$_{*} >$ \mass are nearly all in place by $z \sim 1.5 - 2$ (Conselice
et al. 2007b).  We address this apparent conflict in more detail in \S 5.2.

Although the basic hierarchical model agrees better with the observed
redshift distribution, there are scenarios where the monolithic model can
fit the data as well as the hierarchical scenario (see also Cimatti 
et al. 2002a).   These scenarios require that the K-band
selected galaxies have a significant amount of dust extinction in
the observed K-band.  The amount of extinction needed varies from 1.0 to 0.7
mag at rest-frame $z$ and $R$ bands from $z = 1.5 - 2$ (Kitzbichler \& White
2006).  This extinction needed is even higher at $z = 1 - 1.5$.

By using the colours of galaxies that fit within the $K < 20$ criteria,
we can determine what the contribution of dusty galaxies are to these
counts.  Stellar population synthesis models show that only old
stellar populations, and galaxies with dust extinctions with 
A$_{\rm v} > 1$, have a colour of $(R-K) > 5$ at $z = 1 - 4$.  We can
use this information, and the model results using various dust
extinctions from Nagamine et al. (2005) (Figures~7 and 8), to argue
that these galaxies are not dominated by dust extinction,
which would need to be the case to reconcile the observed K-band
distribution with a monolithic collapse model.

First, the model results compared to data shown in Figures~7 and 8, and 
discussed in \S 4.3 clearly show that K-selected galaxies are not dominated
by heavy dust extinction. At best, only models with E(B-V) = $0 - 0.4$
are able to reproduce the location of real galaxies. The E(B-V) = 0.4
models are however even too red to account for most galaxies in the $(R-K)$
vs. M$_{*}$ diagram (Figure~8).  A representative value of 
R$_{\rm V}$ = A$_{\rm V}$/E(B-V) = 3.1 reveals values of A$_{\rm V} =
 0 - 1.28$ for this range in E(B-V). For the Calzetti et al. (2000)
extinction law, this gives lower values, with A$_{\rm V} = 0 - 0.25$. Thus,
it does not appear likely that our $K < 20$ galaxy sample is dominated by
enough dust extinction (A$_{\rm V} = 1$ needed, on average) to match 
the monolithic collapse galaxy count model.

We find that only 28.3$\pm0.6$\% of $K < 20$ selected galaxies at 
$1 < z < 2$
have a colour $(R-K) > 5$, required to meet the minimum condition for
dust extinction.  We have further argued in \S 4, that a significant
fraction of these EROs are old passively evolving stellar populations,
which are unlikely to have a screen of dust with A$_{\rm V} > 1$
(see e.g., White, Keel \& Conselice 2000). As discussed in \S 4.5.2
over half of the EROs at $1 < z < 2$ appear as early-types, thus
at most only $\sim 14\%$ of the $K < 20$ galaxies at $1 < z < 2$
have a significant amount of dust extinction. The dusty pure-luminosity
evolution monolithic collapse models from Kitzbichler \& White (2006)
predict that all $K < 20$ galaxies have this amount of dust 
extinction, which clearly they do not.  It is thus impossible to
reconcile the K-band redshift distribution with any monolithic model.

In summary our K-band redshift distribution, and those from the
K20 and GOODS surveys, appear to be much closer to the basic
hierarchical model of Kitzbichler \& White (2007) than to
the monolithic collapse model.  We can also rule out pure luminosity
evolution models with significant dust extinction. Therefore, down
to $K = 20$ at $z < 4$ it appears that the monolithic
model cannot be the dominate method for forming galaxies.

\subsection{Structural Evolution}

As discussed in \S 4.5, a large fraction of the EROs we have
HST imaging appear to be undergoing some type of evolution
based on their structural appearance. A logical conclusion is
that many of the peculiar systems are in some phase of a merger
(Conselice et al. 2003a).  We can quantify this by
examining the merger fraction as derived through the CAS
definition of $A > 0.35$ and $A > S$ (Conselice 2003). This
is a strict definition, and will allow us to measure a lower
limit on the merger fraction, even when observing in the
rest-frame ultraviolet (Taylor-Mager et al. 2007).

\begin{figure}
%\vspace{5.5cm}
 \vbox to 80mm{
\includegraphics[angle=0, width=84mm]{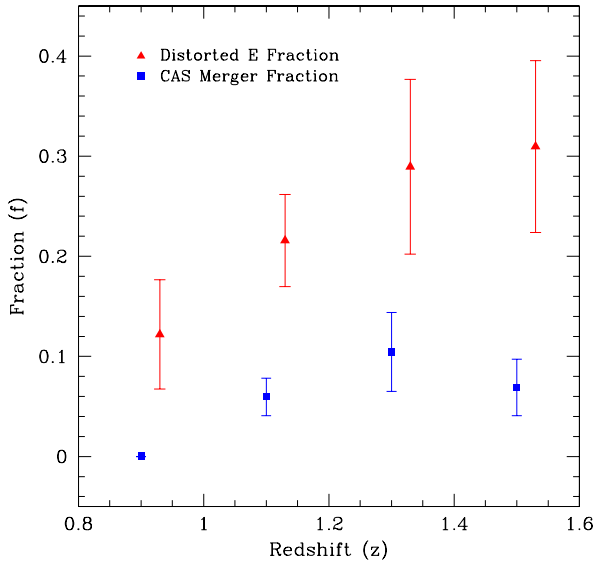}
 \caption{The fraction of galaxies identifiable as a merger
using the CAS system, and the fraction of early-types which
appear to have a distorted structure.}
} \label{sample-figure}
\vspace{2cm}
\end{figure} 

Figure~19 shows the evolution of the merger fraction derived
through this CAS definition. We also include the fraction of
early-types which appear to have a distorted structure.
It appears from this that between $5-10\%$ of the EROs
at $1 < z < 2$ would be identified as a major merger through
the CAS parameters. This is a factor of $\sim 3$ less than the
peculiar fraction (Figure~13), which is what we would expect for a 
population which
is undergoing major mergers (Conselice 2006; Bridge et al. 2007).  The
reason for this is that the CAS approach is sensitive to $\sim
0.4$ Gyr of the merger process (Conselice 2006), while eye-ball
estimates of merging last for $\sim 1.2$ Gyr (Conselice 2006). 
This is another indication
that the peculiar galaxies we are seeing in the ERO population are
in fact undergoing merging. 

Figure~20, as well as Figure~15 demonstrate what evolution is likely
occurring within the ERO population. As can be seen,
the EROs, within our $K < 19.7$ selection, have the same upper range
of masses at $z \sim 0.8 - 2$ (Figure~20). Therefore, little mass 
growth occurs within this
population at this K-band limit.  We also know
that a large fraction of the EROs with M$_{*} >$ \mass are peculiar
in some way. This varies from 64$\pm$24\% at $z > 1.5$ to 41$\pm$6\% at
$1 < z < 1.25$ (\S 4.5).  However, at $z \sim 0$ the fraction of 
M$_{*} >$ \mass red
galaxies which are morphologically early-types is $\sim 85$\% (Conselice
2006a; Conselice et al. 2007b).  

It is clear that a significant
fraction of EROs must have undergone morphological evolution as they cannot
lose mass.   What likely is occurring is that the distorted elliptical
and peculiar galaxies which dominate the population at $z > 1.5$
(Figure~13 and 20) transform into morphologically and spectrally
evolved systems at $z \sim 1$.  The reason this is likely the case
is that there is no difference in the masses of the peculiar and
the early-type EROs.  This also explains why this population, despite
having a mixed morphology clusters so strongly (e.g., Daddi et al. 2000; 
Roch et al. 2003).  Calculations based on the
merger rate in Conselice et al. (2007b) for the most massive
galaxies suggest that on average about one or two major
mergers are occurring for the M$_{*} >$ \mass population at $z < 2$, but
fewer at lower redshifts.

Finally, these major mergers are what may make the K-band counts in the
hierarchical model
redshift distribution higher than the observed.  The reason is that
in the standard hierarchical model, the number and mass densities of
the most massive galaxies with M$_{*} >$ \mass underpredict the observed
number of massive systems by up to two orders of magnitude (Conselice
et al. 2007b).  Within these models the stellar masses for these systems
are largely already formed, but are in distinct galaxies that have
not yet merged (De Lucia et al. 2006).  It is thus easy to see that if
a single massive galaxy at $z \sim 1.5$ was in several pieces, all
of which would still meet the criteria of $K < 20$ based on the
relation of stellar mass and K-mag (Figure~7), then the number of
galaxies with $K < 20$ at higher redshift in the hierarchical model
would be higher than the observed number.
This is consistent with the number densities of massive galaxies being
higher than in the models, as well as for a rapid formation of massive
systems through major mergers at $z > 2$ (Conselice 2006b).

%in /data/papers/eros2/paper2/eroanal
%plot.sm, plot5
\begin{figure}
%\vspace{5.5cm}
 \vbox to 80mm{
\includegraphics[angle=0, width=84mm]{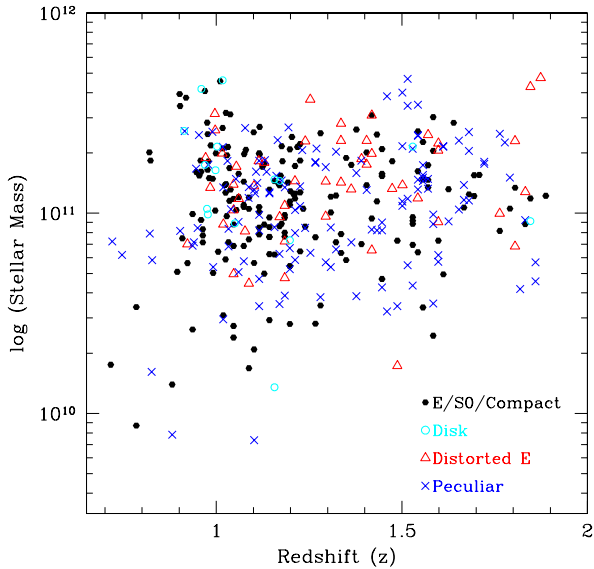}
 \caption{The distribution of ERO galaxy mass with redshift. Labeled
on this figure are the morphological types for each galaxy as determined
through our ACS imaging. Note that massive EROs are detected all
the way out to $z = 2$.}
} \label{sample-figure}
\vspace{2cm}
\end{figure} 

\section{Summary}

In this paper we analyse the faint K-band selected galaxy population as
found in the Palomar NIR survey/DEEP2 spectroscopic survey overlap.
Our primary goal is to determine the nature of the faint $K > 19$ 
galaxy population.  While many of these galaxies are too faint
for detailed spectroscopy, we can investigate their nature through 
spectroscopic and photometric
redshifts, stellar masses, as well as photometric and structural features. 
Our major findings include: \\

\noindent 1. The redshift distribution for K-selected galaxies depends
strongly on apparent K-magnitude. Most systems at 
$K < 17$ are at $z < 1.4$,
while a significant fraction of sources with $17 < K < 19$ are at $z > 2$.
These K-bright high$-z$ galaxies are the progenitors of today's massive
galaxies.\\

\noindent 2. We find that a significant fraction (28.3$\pm0.6$\%)
of the $K < 20$ galaxy
population consists of extremely red or massive galaxies at $z > 1$. We
characterise the population of log M$_{*} > 11$ sources in 
Conselice et al. (2007b), while we analyse the extremely red objects
(EROs) in this paper.\\

\noindent 3. We find that EROs at $K < 19.7$ are a
well defined population in terms of redshifts and masses.  Nearly all EROs
are at $z > 1$, and have stellar masses with M$_{*} > 10^{11}$ \solm.  EROs
are therefore certainly the progenitors of today's massive galaxies.
The corollary to this however is  not necessarily true.  There are
massive galaxies at $z > 1$ that would not be selected with the ERO
criteria.  We find that the ERO selection locates 35-75\% of all
ultra-massive, M$_{*} > 10^{11.5}$ \solm galaxies, at $z = 1-2$, while
only 25\% of \mass $<$ M$_{*} < 10^{11.5}$ \solm galaxies are located 
with this  colour cut.\\

\noindent 4. We examine the morphological and structural properties
of our ERO sample and find, as others previously have, a mixed population
of ellipticals and peculiars. In total, we find that the ERO population
is dominated by early-type galaxies, with an overall fraction of 57$\pm3$\%
of the total.  Interestingly, we find that a significant
fraction of the early-types ($\sim 25$\%) are distorted ellipticals, which
could be classified as peculiars, although these systems at a slightly
lower resolution than ACS, or using a quantitative approach would be
seen as early-types.   Peculiars account for the remaining 34\%, and
many of these are likely in some merger phase.  This fraction tends
to evolve such that the peculiars are the dominant population at
higher redshifts, $z > 1$.\\

\noindent 5. We investigate the structural parameters for our EROs
using the CAS system.  We find that visual estimates of galaxy class
and position in CAS space roughly agree, although the asymmetries
of these systems are higher than what their visual morphologies
would suggest.  We find a bimodality in the stellar mass-concentration
diagram where the peculiar EROs are at a low concentration and the
early-types are highly concentrated.  The distorted ellipticals
fall in between these two populations suggesting an evolutionary
connection in the passive evolution of galaxy structure within
the ERO population.  \\

\noindent 6. We compare the $(R-K) > 5.3$ ERO selection with the
$(I-K) > 4$ ERO selection, and find that the $(I-K) > 4$ ERO are at 
slightly higher redshifts than the $(R-K) > 5.3$ selection,
suggesting that it is a more useful criteria for finding evolved
galaxies at $z > 1.5$.\\

\noindent 7.  By examining the redshift distribution of $K < 20$
galaxies, and comparing to monolithic collapse and hierarchical formation
models, we are able to rule out all monolithic collapse models for
the formation of massive galaxies.  These monolithic collapse models
predict a higher number of $K < 20$ galaxies as a function of redshift
at a significance of $\sim 10$ $\sigma$.  While some monolithic collapse
models are able to reproduce our galaxy counts, these are dominated
by dust. We are however able to show that only $\sim 14$\% of the $K < 20$
galaxies have potentially enough dust to match this model, the others
being evolved galaxies, or blue star forming systems. \\

The Palomar and DEEP2 surveys would not have been completed without the active
help of the staff at the Palomar and Keck observatories.  We particularly
thank Richard Ellis and Sandy Faber for advice and their participation
in these surveys. 
We thank Ken Nagamine and Manfred Kitzbichler for providing their
models and comments on this paper. We also thank the referee for their
careful reading and commenting on this paper.  We acknowledge
funding to support this effort from a National Science Foundation
Astronomy \& Astrophysics Fellowship, grants from the UK Particle
Physics and Astronomy Research Council (PPARC),
Support for the ACS imaging of the EGS in GO program 10134 was 
provided by NASA through NASA grant HST-G0-10134.13-A from the Space Telescope Science Institute, which is operated by the Association of Universities for Research in Astronomy, Inc., under NASA contract NAS 5-26555. 
 JAN is supported by NASA through Hubble Fellowship grant HF-01182.01-A/HF-011065.01-A.
The authors also wish to recognise and acknowledge the very significant cultural role and reverence that the summit of Mauna Kea has always had within the indigenous Hawaiian community. We are most fortunate to have the opportunity to conduct observations from this mountain.

\label{lastpage}

\end{document}